# Light-induced anomalous Hall conductivity in massive 3D Dirac semimetal $Co_3Sn_2S_2$

**Authors:**

Naotaka Yoshikawa[1*], Shun Okumura[2,3], Yoshua Hirai[1], Kazuma Ogawa[1], Kohei Fujiwara[4], Junya Ikeda[4], Akihiro Ozawa[2], Takashi Koretsune[5], Ryotaro Arita[1,6], Aditi Mitra[7], Atsushi Tsukazaki[4,8], Takashi Oka[2], Ryo Shimano[1,9*]

**Affiliations:**

[1] Department of Physics, The University of Tokyo, Hongo, Tokyo 113-0033, Japan.

[2] Institute for Solid State Physics, The University of Tokyo, Kashiwa 277-8581, Japan

[3] Department of Applied Physics, The University of Tokyo, Hongo, Tokyo 113-8656, Japan

[4] Institute for Materials Research, Tohoku University, Sendai 980-8577, Japan.

[5] Department of Physics, Tohoku University, Miyagi 980-8578, Japan

[6] RIKEN Center for Emergent Matter Science, 2-1 Hirosawa, Wako 351-0198, Japan

[7] Center for Quantum Phenomena, Department of Physics, New York University, 726 Broadway, New York, NY, 10003, USA

[8] Center for Science and Innovation in Spintronics (CSIS), Core Research Cluster, Tohoku University, Sendai 980-8577, Japan.

[9] Cryogenic Research Center, The University of Tokyo, Yayoi, Tokyo 113-0032, Japan.

[*]Correspondence to: yoshikawa@thz.phys.s.u-tokyo.ac.jp, shimano@phys.s.u-tokyo.ac.jp

**Abstract**

Weyl semimetals can emerge from Dirac semimetals when the time-reversal or spatial-inversion symmetries are broken. Recently, it has been proposed based on the Floquet theory that Dirac semimetals can be converted into Weyl semimetals even by shining circularly polarized light (CPL). Here we have investigated the possibility of such a Dirac-Weyl conversion by measuring the CPL-induced anomalous Hall conductivity (AHC) in a massive 3D Dirac semimetal $Co_3Sn_2S_2$ in the paramagnetic phase using ultrafast mid-infrared pump-terahertz Faraday rotation probe spectroscopy. We find that the field-strength and driving frequency dependence of the observed AHC is well accounted for by CPL-induced nonzero Berry curvature associated with the splitting of the Dirac bands as predicted by the Floquet theory. The estimated splitting of the Dirac bands reaches about 60 % of the mass gap and the calculated CPL-induced AHC quantitatively reproduces the experimental observation, demonstrating a promising route toward the realization of Floquet-Weyl states from massive Dirac semimetals.

## Main text

## I. INTRODUCTION

Three-dimensional (3D) Dirac semimetals (DSMs) and Weyl semimetals (WSMs) have gained a growing interest because of the presence of massless quasiparticles around the linearly crossing electronic bands, that effectively act as Dirac and Weyl fermions, respectively [1–7]. These quantum materials exhibit various intriguing physical properties such as large linear magnetoresistance [8], chiral-anomaly-induced negative magnetoresistance [9,10], surface Fermi arc states [11], unusual quantum oscillations [12], and intrinsic anomalous Hall effect (AHE) [13]. Such electronic functionalities mostly associated with topological properties have made DSMs and WSMs highly attractive from fundamental and practical points of view.

Today, light is considered an ideal tool for manipulating quantum electronic states of matter in an ultrafast manner. Ultrafast dynamics of electrons in DSMs and WSMs have been intensely investigated recently, revealing novel optoelectronic phenomena such as the circular photogalvanic and photovoltaic effects [14–17], which are related to the coupling between the light polarization and the pseudospin of the massless fermions. Furthermore, transient modifications of the crystal symmetry characterizing the inversion-symmetry breaking WSM states by light have been reported as a route to the dynamically induced topological phase transition via the phonon excitation or photocurrent generation [18–20].

Distinct from these processes accompanying the optical excitation of quasiparticles or collective modes, another route to manipulate quantum materials is to change the electronic states and band topology using their coupling to laser fields, dubbed as Floquet engineering [21–23]. Theoretically, Floquet engineering of two-dimensional (2D) systems has led to proposals of the Floquet topological insulators [24,25], while in 3D a conversion from 3D DSMs to WSMs by CPL driving, namely, Floquet-WSM has been proposed [26–28]. A time-periodic CPL field in massless 3D DSMs generates a chiral gauge field (CGF), which induces a momentum shift that is positive (negative) for left-handed (right-handed) chirality state, and thus, the spin-degenerate Dirac cones will split into Weyl cones. Here, the separation of the Floquet-Weyl points (WPs) in the momentum space corresponds to twice of the CGF $\boldsymbol{A_5} = \beta\hat{\boldsymbol{z}}$ perpendicular to the polarization plane of irradiated CPL with its magnitude scaling as $\beta \propto E^2/\Omega^3$ ($E$: electric field strength, $\Omega$: angular frequency of the driving field). The light-induced CGF can be tuned by the direction and intensity of light, leading to the manipulation of emergent topological functionalities such as AHE and to versatile topological phases which only appear in the nonequilibrium state [28–30]. In the case of massive 3D DSMs, CPL driving causes the original Dirac bands to split proportional to $\beta$, eventually leading to the Floquet-Weyl states when $2\beta$ exceeds the original Dirac gap [27,28]. This highlights another intriguing aspect of the Floquet-WSM state, namely, one can create topologically protected massless fermions even in massive DSMs.

Experimentally, the Floquet-Bloch states were observed in the surface state of topological insulators [31,32] and semiconductors [33,34], and modulated optical responses due to the Floquet states were demonstrated under below-gap excitation [35,36]. These successful demonstrations mainly rely on the formation of photon sidebands and level repulsions between Floquet states with different photon numbers. Distinct from those responses arising from high energy band structures at the photon resonances, one intriguing aspect of the Floquet engineering is that it can manipulate the original bands in the low-energy region, i.e. the 0-photon states, through the implementation of topological gap in 2D DSMs and CGF in 3D DSMs. To date, the experimental demonstration in this regard is limited to the observation of a gap at the 2D Dirac point (DP) in the topological insulator under the CPL driving [31]. Although the CPL-induced AHE in graphene has been observed [37], various origins are argued to give rise to the anomalous Hall conductivity (AHC) and the discrimination of those contributions remains under debate [38]. The demonstration of the original band engineering by forming the Floquet states in 3D systems and its applicability to versatile materials have remained to be explored.

In this work, we investigated a Dirac-Weyl conversion in a massive 3D DSM $Co_3Sn_2S_2$ driven by the CPL laser pulses. The anomalous Hall response of the driven $Co_3Sn_2S_2$ was measured in a time-resolved manner, of which the dynamics, helicity dependence and most importantly the scaling on the electric field strength and driving frequency are well explained by CPL-induced nonzero Berry curvature associated with the splitting of the Dirac bands as predicted by the Floquet theory.

**II. SETUP**

To investigate the Floquet states emergent in a 3D DSM, we focus on the Kagome-lattice shandite $Co_3Sn_2S_2$ (Fig. 1(a)). $Co_3Sn_2S_2$ has recently attracted considerable attention due to the emergence of the ferromagnetic WSM phase below the Curie temperature $T_C$ around 175 K [39,40], while it also provides a playground for investigating 3D Dirac electrons in its paramagnetic phase above $T_C$ [41,42]. In the ferromagnetic phase, $Co_3Sn_2S_2$ shows three pairs of Weyl nodes with linearly dispersing bands in the 3D momentum space, which arises from band inversion, exchange splitting, and spin-orbit coupling (SOC). Without SOC, the valence and conduction bands invert near the $L$ points of the Brillouin zone, which forms nodal rings protected by mirror symmetry. The spin band degeneracy is lifted by the exchange coupling in the ferromagnetic phase. The nodal rings are gapped out by including SOC, except for three pairs of linearly crossing points which give rise to Weyl nodes [39,40].

In the paramagnetic phase above $T_C$, first-principles calculations reveal that the original nodal rings are fully gapped out by including SOC as shown in Fig. 1(b) (for the detail see Appendix A) [41,43]. The energy difference between the two bands has a minimum value of about 10 meV, around which the band structure can be described by a massive 3D DSM.

In this study, we performed mid-infrared (MIR) pump-terahertz (THz) Faraday rotation probe spectroscopy as illustrated in Fig. 1(c) (for the detailed setup see Appendix B). A MIR femtosecond pulse was used as a periodic driving field to induce the Floquet-Weyl state without inducing material damage. The effect of band structure modification induced by the time-reversal symmetry breaking CPL field is probed through the AHC, which can be detected by observing the THz Faraday rotation signal. This is because the Faraday effect in the low energy (~THz) range represents the ac-Hall conductivity and can be recognized as ac- or optical-Hall effect [44–48]. In fact, in the present system of $Co_3Sn_2S_2$, the THz Faraday rotation has been proven to reflect the intrinsic dc AHE [49,50]. Importantly, the measurement of the AHC using a THz pulse can be used as an ultrafast probe of the ac-AHE. As a sample, we used a c-axis oriented, 23 nm-thick $Co_3Sn_2S_2$ film grown on a sapphire substrate [51] (See Appendix C for details). All the measurements were performed at room temperature.

## III. RESULT

### 1. Light-induced anomalous Hall conductivity

Figure 1(d) presents time evolutions of the THz Faraday rotation of the $Co_3Sn_2S_2$ film induced by the MIR pump pulse for right-hand circular polarization ($\sigma^+$, red circles) and left-hand circular polarization ($\sigma^-$, blue circles). The center frequency of the MIR pump was 75 THz. The fluence and pulse width is measured as $2.4\ \mathrm{mJ\ cm^{-2}}$ and 160 fs in FWHM, respectively. Using the measured optical conductivity (data shown in Appendix D), the peak electric field inside the film sample is estimated as $E = 1.0\ \mathrm{MV\ cm^{-1}}$. A transient Faraday signal is observed that reverses sign upon reversing the helicity of the pump CPL. Gray line in Fig. 1(d) indicates the cross-correlation waveform between the pump MIR pulse and the near-infrared pulse used for the optical sampling of the probe THz pulse. The temporal profile of the observed Faraday rotation overlaps this cross-correlation waveform, indicating that the signal arises only during the pump pulse irradiation.

The spectra of complex Faraday rotation angle, $\Delta\Theta(\omega) = \Delta\theta(\omega) + i\Delta\eta(\omega)$ can be obtained by measuring the THz waveforms with parallel ($E_x$) and perpendicular ($E_y$) to the incident polarization [46–48]. The typical spectrum is displayed in Fig. 2(a) which was measured at the fixed delay $t_{\mathrm{pp}} = 0$ in Fig. 1(d). The CPL-induced AHC $\Delta\sigma_{xy}(\omega)$ is calculated from the Faraday rotation angle through the equation $\Delta\Theta(\omega) \cong Z_0\Delta\sigma_{xy}(\omega)d/[1 + n_{\mathrm{sub}} + Z_0\sigma_{xx}(\omega)d]$ for a thin-film sample, where $Z_0$, $d$, $n_{\mathrm{sub}}$ and $\sigma_{xx}(\omega)$ are the vacuum impedance, the thickness of the film, the refractive index of the substrate and the longitudinal optical conductivity of the film sample, respectively. Plotted in Fig. 2(b) and (c) are the real and imaginary part of the AHC spectrum $\Delta\sigma_{xy}(\omega)$ with various pump fluences. The real part ($\mathrm{Re}\,\Delta\sigma_{xy}(\omega)$) is spectrally flat, and its magnitude monotonically increases on increasing the fluence. The imaginary part of the Hall conductivity ($\mathrm{Im}\,\Delta\sigma_{xy}(\omega)$) is close to zero and virtually unchanged as the fluence increases. We also numerically confirmed that the measured nonequilibrium conductivity indeed reflects the dynamics induced by the MIR pump pulse in the temporal region shorter than the duration of the probe THz pulse (Appendix E).

### 2. Field-strength and driving frequency dependence

To elucidate the origin of the CPL-induced AHC, we examined the pump fluence and frequency dependences. For this purpose, we used several frequencies varying from 50 to 120 THz and measured the CPL-induced AHC (for the characterization of the pump pulse, see Appendix B). Figure 3(a)-(f) present the CPL-induced AHC for the various driving frequencies as a function of the squared peak electric field inside the $Co_3Sn_2S_2$ film. Here, we plotted the average values of $\mathrm{Re}\,\Delta\sigma_{xy}(\omega)$ between 6.2 and 8.3 meV (we chose this range by considering the spatial uniformity of the pump beam, for details see Appendix B). The field-strength dependence shows the quadratic behavior $\Delta\sigma_{xy} \propto E^2$ for all the driving frequencies in the weak excitation regime and shows a deviation from $E^2$ dependence at higher excitation intensity. Then, the efficiency of the CPL-induced AHC is evaluated by line-fitting using the data in the weak excitation regime as indicated by solid lines in Fig. 3(a)-(f). Figure 3(g) plots the obtained frequency dependence of the AHC efficiency. The result is well fitted by a power law of $\Omega^{-2.89\pm0.29}$ (blue line), namely the CPL-induced AHC is universally scaled by $\Delta\sigma_{xy} \propto E^2/\Omega^3$ over a wide range of driving frequency. This result agrees well with the picture predicted by the Floquet theory, namely $\Delta\sigma_{xy}$ is proportional to the magnitude of the CPL-induced CGF, $\beta = |\boldsymbol{A}_5| \propto E^2/\Omega^3$, for massless 3D DSMs, which also applies to the case of massive DSMs as we describe in detail in the next section.

We also measured the driving frequency dependence of the CPL-induced change of the longitudinal optical conductivity, $\Delta\sigma_{xx}(\omega)$ (for data set and detailed discussion, see Appendix F), which shows a long-lived, helicity-independent, and nearly frequency-independent behavior, making a stark contrast to that of $\Delta\sigma_{xy}(\omega)$. This is rather reasonable because $\Delta\sigma_{xx}(\omega)$ is attributed to the pump-induced change of carrier density and/or scattering rate of the metallic carriers.

### 3. Effective continuum model for the Floquet state of $Co_3Sn_2S_2$

Here, to qualitatively understand the experimental results, we theoretically developed an effective continuum model to describe the band structure of $Co_3Sn_2S_2$ based on the results of first-principles calculations (for the details see Appendix A). Considering the electronic state of $Co_3Sn_2S_2$ noted in Sec. II, we construct a 4 × 4 Hamiltonian extracting the spin and two orbital degrees of freedom, the valence and conduction bands, mainly from the *d* orbitals of Co atoms. The non-driven state with a pair of DPs can be modeled by

$$H(\mathbf{k}) = H_{\mathrm{NR}}(\mathbf{k}) + H_{\mathrm{SO},z}(\mathbf{k}) + H_{\mathrm{SO},x}\,, \qquad (1)$$

where $H_{\mathrm{NR}}(\mathbf{k}) = [M - Bk^2]\tau_z s_0 + v k_x \tau_x s_0$ describes a nodal-ring semimetal and the SOC associated with the out-of-plane spin, $H_{\mathrm{SO},z}(\mathbf{k}) = \Delta_{\mathrm{SO}} k_y \tau_y s_z$, gaps it out leaving two Dirac nodes. Another in-plane type of SOC, $H_{\mathrm{SO},x} = m\tau_y s_x$ gaps out the DPs with the gap energy of $2m$. Here, we assume $\hbar = e = 1$. $M$ and $B$ are positive constants, $v$ refers to the group velocity at the DPs along $x$-direction, and $\Delta_{\mathrm{SO}}$ quantifies the out-of-plane type of SOC. $\tau_{x,y,z}$ ($s_{x,y,z}$) and $\tau_0$ ($s_0$) are Pauli matrices and the identity matrix for the orbital (spin) quantum number, respectively. The calculated band structure is plotted in Fig. 4(a), along with the enlarged view around the massive DP in the inset. The time-dependent Hamiltonian $H(t)$ with right-hand CPL along the $z$-direction $\mathbf{A}(t) = \left(\frac{E}{\Omega}\cos(\Omega t), -\frac{E}{\Omega}\sin(\Omega t), 0\right)$ is given by applying Peierls substitution as $H(t) = H\big(\mathbf{k} + e\mathbf{A}(t)\big)$. We obtain the Floquet Hamiltonian $H_n = \frac{1}{T}\int_0^T dt\, H(t) e^{in\Omega t}$ by Fourier transformation for time $t$, where $T = \frac{2\pi}{\Omega}$. In order to discuss the effect of original band modulation, we obtain the effective Floquet Hamiltonian by high-frequency expansion [21,23,30,52];

$$H_{\mathrm{eff}}(\mathbf{k}) \approx H_0 + \sum_{n\geq 1} \frac{[H_{-n}, H_{+n}]}{n\Omega}, \qquad (2)$$

where $H_0 = [\widetilde{M} - Bk^2]\tau_z s_0 + v k_x \tau_x s_0 + \Delta_{\mathrm{SO}} k_y \tau_y s_z + m\tau_y s_x$, $\frac{[H_{-1}, H_{+1}]}{\Omega} = \beta\tau_z s_z + \frac{2\beta B}{v} k_x \tau_x s_z + \frac{2\beta B}{\Delta_{\mathrm{SO}}} k_y \tau_y s_0$, and $\frac{[H_{-n}, H_{+n}]}{n\Omega} = 0$ for $n \geq 2$. In $H_0$, the model parameter $M$ is renormalized by irradiation of light as $\widetilde{M} = M - \frac{E^2 B}{\Omega^2}$. It is worth noting that $\beta = \frac{E^2 v \Delta_{\mathrm{SO}}}{\Omega^3}$ in the first term of $\frac{[H_{-1}, H_{+1}]}{\Omega}$ is a leading term of the CPL-induced band splitting of the Dirac bands resulting in the nonzero Berry curvature.

By diagonalizing this $4 \times 4$ effective Floquet Hamiltonian, the Floquet band structure is obtained as drawn in Fig. 4(b) and (c) for $\beta < m$ and $\beta > m$, respectively. Since the electronic states around the band crossings in this model can be approximately described by Dirac Hamiltonian, the effect of the induced band modification to these states is essentially same as that to ideal 3D DSMs [27,28]. For the case of massive 3D DSMs, in the weak field region described by $\beta < m$, the degenerated Dirac bands split while the mass gap remains (Fig. 4(b)). With increasing the field strength so that the condition $\beta > m$ is realized, two pairs of Floquet-Weyl nodes emerge with the separation given by $\sqrt{\beta^2 - m^2}$ (Fig. 4(c)). It should be noted here that, for the case of massless 3D DSMs, this $\beta$ corresponds to the magnitude of the CGF $\boldsymbol{A}_5 = \beta\hat{\boldsymbol{z}}$ which is equivalent to the momentum separation between the WPs with opposite chirality. In massive 3D DSMs, the chirality is no longer conserved so CGF cannot be defined for $\beta < m$, but the infinitesimal $\beta$ induces the splitting of the Dirac bands which is asymptotic to a momentum shift vector $\boldsymbol{A_5}$ for the electrons away from the mass gap.

The AHC of the Floquet state of the developed continuum model was calculated from linear response theory [24,53] (see Appendix G for details). For this calculation we assume the Fermi-Dirac distribution of the electrons in the Floquet state with the chemical potential (0.12 eV below the center of the mass gap) and temperature (300 K) relevant to the experiment. The calculated value is multiplied by 3 considering there are three pairs of DPs in the first Brillouin zone of $Co_3Sn_2S_2$. We also consider the anisotropy of the nodal ring and the propagation direction of the CPL tilting from the separation of the DPs, which do not change the essential nature but affects the value of AHC (for details see Appendix G). The calculated spectra displayed in Fig. 4(d) and (e) capture the features observed in the experiment (Fig. 2), where the real part shows a constant value while the imaginary part is nearly zero. The field-strength dependence and frequency dependence of the AHC, $\mathrm{Re}\,\Delta\sigma_{xy}(\hbar\omega = 10\ \mathrm{meV})$, are presented in Fig. 4(f) and (g), clearly indicating $\mathrm{Re}\,\Delta\sigma_{xy} \propto E^2\Omega^{-3} \propto \beta$. This result ensures that the CPL-induced band structure modification contributes to the AHC even if the Fermi energy is away from the Dirac nodes and the energy bands at the Fermi energy have finite curvatures. This is because the energy states around the Fermi energy are split by the amount proportional to $\beta$ in energy, causing the nonzero value of the integrated Berry curvature over the occupied states. It should also be noted that a finite AHC proportional to $\beta$ is observed even when the field is not strong enough to close the mass gap ($\beta < m$). This is reasonable because the AHC mainly originates from the energy band splitting around the Fermi energy which is away from the Dirac mass gap in the present case. Indeed, the maximum $\beta$ in the present experiment is estimated as 3.1 meV for $\Omega/2\pi = 55\ \mathrm{THz}$ and $E = 0.69\ \mathrm{MV\ cm^{-1}}$, which is about 60 % of the amount required for closing the mass gap. The amount is given by $\beta = (eE)^2 v_x v_y/(\hbar\Omega)^3$ ($v_{x(y)}$ are the group velocities of the Dirac cone) under CPL with its polarization in $xy$-plane (for the estimation of the group velocities around the DP of $Co_3Sn_2S_2$, see Appendix A). Thus, the experimental observation of the CPL-induced AHC proportional to $\beta$ is reasonably explained by the CPL-induced modification of the Dirac bands accompanied by the nonzero integrated Berry curvature, which will evolve into the Floquet-Weyl state when $\beta$ exceeds $m$.

## 4. Effective tight-binding model for the Floquet state of $Co_3Sn_2S_2$

To mimic the multiband structure, we also investigated the Floquet state of $Co_3Sn_2S_2$ by the tight-binding model on the layered Kagome lattice. The effective tight-binding Hamiltonian for the paramagnetic phase of $Co_3Sn_2S_2$ has been developed by extracting each one *d* orbital from three Co atoms on the Kagome layer and one *p* orbital from interlayer Sn atom [54–56], which is described by $H_{\mathrm{TB}}(\mathbf{k}) = H_{\mathrm{dp}}(\mathbf{k}) + H_{\mathrm{KM}}(\mathbf{k}) + H_{\mathrm{sR}}(\mathbf{k})$. Here, $H_{\mathrm{dp}}(\boldsymbol{k})$ denotes the spin-independent hopping term which forms the nodal rings, $H_{\mathrm{KM}}(\mathbf{k})$ is the Kane-Mele type SOC which gaps out the nodal ring except for the DPs, and $H_{\mathrm{sR}}(\mathbf{k})$ is the staggered Rashba type SOC which further opens small gaps at the DPs (for more detail, see Appendix H). Their roles are similar to the terms in the continuum model in Eq. (1) in order.

The calculated non-driven band structure is plotted as black dashed lines in Fig. 5(a). The time-independent effective Hamiltonian under CPL driving is calculated by applying high-frequency expansion in the similar procedure for the continuum model, namely, Eq. (2). The obtained band structures are indicated by turquoise ($E = 1.0\ \mathrm{MV\ cm^{-1}}$) and orange ($E = 3.2\ \mathrm{MV\ cm^{-1}}$) lines in Fig. 5(a). The inset shows an enlarged view where band shifts and splitting are identified even at $E = 1.0\ \mathrm{MV\ cm^{-1}}$ which is adjusted to our experiment. At $E = 3.2\ \mathrm{MV\ cm^{-1}}$, these features become more prominent. The band structure around the massive DP, which is not located at the momentum connecting the high-symmetry points, are displayed in Fig. 5(b). Formation of WPs is identified for the strong CPL driving with $E = 3.2\ \mathrm{MV\ cm^{-1}}$, while band splitting is induced even for the weak driving with $E = 1.0\ \mathrm{MV\ cm^{-1}}$. Due to the band splitting of the positive and negative Berry curvature states (see Fig. 15(a) in Appendix H for the Berry curvature distribution in the band structure), a finite difference arises in the positive and negative densities of Berry curvature, though they still appear nearly identical, as shown in Fig. 5(c). Then, the AHC spectrum of the Floquet state is calculated by Eq. G4 with the assumption of Fermi-Dirac distribution, which is presented in Fig. 5(d). For the photon energy below 0.01 eV that corresponds to our experimental condition, the real part is constant and the imaginary part is almost zero, showing a good agreement with the experimental result (Fig. 2(b) and (c)). The absence of spectral peaks in the THz frequency range relevant to the band splitting can be attributed to the fact that the splitting energy is smaller than the damping (0.01 eV) which was incorporated in the linear response calculation. The validity of the damping constant will be shown later in the discussion section. A peak structure that appears at above 0.1 eV in the calculated AHC spectrum is attributed to interband transitions in the Floquet 0-photon states. The field strength and driving frequency dependence of the CPL-induced AHC, $\mathrm{Re}\,\Delta\sigma_{xy}(\hbar\omega = 10\ \mathrm{meV})$, is plotted in Fig. 5(e) and (f), respectively. The tight-binding model reproduces the complex multiband structure with the Fermi surfaces far from those of the minimal model of 3D DSMs, nevertheless the calculation demonstrates the scaling law of Re $\Delta\sigma_{xy} \propto E^2\Omega^{-3}$, in accordance with the experimental observation. Here we assume the Fermi-Dirac distribution of the electrons again, and calculated AHC with the electron temperature of 300 K and 1000 K. Since $\sigma_{xy}$ is determined by the Berry curvature integrated over the entire occupied states, there is no universal trend such as being larger or smaller at higher temperature. Importantly, the calculation for the two temperatures indicates that the increase in the electron temperature, which is expected to occur due to the photoexcitation, does not significantly suppress $\sigma_{xy}$. The calculations here also demonstrate that even for the situation where the band splitting (6 meV at DP) is smaller than the energy scale of the thermal distribution ($k_{\mathrm{B}}T$ =26 meV for 300 K) and the damping, the AHC of the Floquet state well reproduces the experimental observation. The quantitative agreement between the calculation and the experiment (see Appendix H for more discussion from the quantitative perspective), further supports the interpretation that the band structure modification predicted by the Floquet theory is realized.

## IV. DISCUSSION

As a possible contribution of the CPL-induced AHC, here we consider the photon resonance effect, i.e., the contributions from Berry curvatures around the level crossing between the original bands and Floquet sidebands at the resonant transition [38,57,58]. In particular, the one-photon resonance with SOC can cause a Floquet-Weyl state at the resonant momentum in a wide range of 3D materials, which should also contribute to AHC [57]. Such a resonant contribution scales as $\Delta\sigma_{xy} \propto E$ in the strong-field regime, but in the weak-field regime it scales as $E^2$ which itself does not contradict with the present experimental result. In Fig. 3(a)-(f), the deviation of $\Delta\sigma_{xy}$ from $E^2$ dependence is prominent at the driving frequency of 50 and 73 THz. The field strength at which the deviation is identified is not constant nor monotonic with changing the driving frequency. Considering this fact, the deviation from $E^2$ may stem from the complex photon resonances in $Co_3Sn_2S_2$ in the MIR range. Yet, the universal scaling law of AHC ($\Delta\sigma_{xy} \propto E^2\Omega^{-3}$) in the weak field regime is hardly attributed to such a resonant excitation contribution. To theoretically investigate the resonant contribution in $Co_3Sn_2S_2$, we calculated the Floquet quasi-energy band structure with including the multiple photon sectors (see Appendix I). The AHC exhibits a significantly nonmonotonic dependence on the driving frequency (Fig. 16(b)), which is likely attributed to the resonant contributions to the Berry curvature arising from the hybridization of the Floquet bands with different photon numbers. Furthermore, the calculations revealed that the balance between the contributions from the Floquet 0-photon state and the sidebands strongly depends on the damping constant which is caused by scattering processes (Fig. 16(c)). As the damping constant increases, the resonant contribution is significantly suppressed, while the 0-photon contribution shows much weaker dependence. This can be reasonably understood by considering that the effect of band hybridization strongly reflects the deviation of the electron distribution from the equilibrium state and tends to be suppressed in systems with significant scattering which promotes the depopulation and broadening of Floquet replica bands. Therefore, we infer that the substantial electron scatterings in $Co_3Sn_2S_2$ would suppress the resonant contributions to the CPL-induced AHC, making the contribution of the Floquet 0-photon state dominant. Indeed, we estimate the transport scattering rate of our $Co_3Sn_2S_2$ sample to be $\hbar\gamma_{\mathrm{tr}} = 81 \pm 9$ meV from the measured optical conductivity (Appendix D), which is larger than the estimated Rabi energy, $\Delta_{\mathrm{Rabi}} = eEv\Omega^{-1} = 28.5$ meV (for $\Omega/2\pi = 55$ THz, $E = 0.69$ MV cm$^{-1}$, and $v = 0.85$ eV $\cdot$ Å) as a typical hybridization energy. We also note that the scattering rate in $Co_3Sn_2S_2$ is larger than that of other Dirac materials, for example, graphene [37] and $Cd_3As_2$ [59], which may account for the difference in the relative balance between the 0-photon and resonant contributions compared to those materials.

Scattering processes generally inhibit the coherent motion of electrons induced by light field, and thus how robust the Floquet states are generated under the presence of dissipation and decoherence has been extensively investigated theoretically [38,60–63]. For instance, the robustness of the Floquet states of graphene under the scattering rate of about one order smaller than the driving photon energy has been theoretically demonstrated, where the scattering processes are shown to cause mainly the population redistribution and broadening of the Floquet bands [62]. The robustness of Floquet states has also been theoretically discussed in terms of Floquet fidelity [38,60]. It has been shown that, Floquet states can be established in graphene throughout a large portion of the Brillouin zone except for the one-photon resonance despite the presence of dissipation. In the present case of $Co_3Sn_2S_2$, the scattering rate (81 meV) is at most about one-third of the driving photon energy $\hbar\Omega = 0.21 - 0.50$ eV used in this study. Our experimental observation of $\Delta\sigma_{xy} \propto E^2\Omega^{-3}$ suggests that the response can be described by the original (0-photon) bands of the Floquet state even under such substantial scatterings. This result does not contradict with the recent results concerning the robustness of the Floquet bands as described above.

Another possible origin of CPL-induced AHC is the population imbalance effect, where the interband transition under the static electric field (in our case the electric field of the probe THz pulse) causes an asymmetric population in the momentum space perpendicular to the applied electric field, leading to a Hall current. This was proposed as a dominant process for the experimentally observed CPL-induced AHC in graphene [38,60] and $Cd_3As_2$, a 3D DSM [59]. However, we infer that this mechanism is also strongly influenced by the nonequilibrium electron distribution and suppressed in materials with strong relaxation. Furthermore, it is hard to ascribe the observed frequency scaling to this population imbalance effect as we discuss in the following. The Hall current of this mechanism is given by **k** integration of the product of the group velocity along the transverse direction and the population imbalance that is derived from circular dichroism induced by the dc electric field. For an ideal isotropic 3D DSM, namely, in the case that the one-photon transition occurs within linearly dispersing Dirac cone with the constant group velocity, the driving frequency dependence of the Hall current associated the population imbalance effect can be calculated to show $\propto E^2\Omega^{-2}$ [59]. This power-law is not consistent with the experimentally observed one in this study, $\propto E^2\Omega^{-3}$, whereas the frequency scaling may change if the Dirac bands are no more relevant to the one-photon absorption. In the present case, $\hbar\Omega$ is substantially larger than the width of the linearly dispersing bands, and one-photon transition will occur at the various momenta due to the complicated band structure of $Co_3Sn_2S_2$. In such a case, both the group velocity and the population imbalance should show a complex dependence on the driving frequency. Therefore, it is unlikely that the observed CPL-induced AHC exhibiting the universal scaling law of $\propto E^2\Omega^{-3}$ over a wide frequency range originates from the population imbalance effect.

Lastly, we consider the CPL-induced Zeeman term in the Floquet state of $Co_3Sn_2S_2$, which we refer to as inverse Faraday effect (IFE). In addition to the band modification equivalent to CGF for massless 3D DSMs, the Zeeman field can be implemented in the CPL-driven Floquet states via the in-plane type SOC which also scales as $\propto E^2\Omega^{-3}$ (for more discussion, see Appendix I). To evaluate the contribution of the IFE to the AHC in $Co_3Sn_2S_2$, we calculated the CPL-induced AHC with and without staggered Rashba type SOC, $H_{\mathrm{sR}}(\boldsymbol{k})$, which is the in-plane type SOC in the effective tight-binding model. The calculation in Fig. 16(c) in Appendix I shows that the effect of including the staggered Rashba type SOC is minor (~7 %) to the CPL-induced AHC, indicating that the CGF-like band splitting due to the unique coupling between Dirac electrons and light provides the dominant contribution to the observed AHC in the present case. The stark contrast in the relative magnitudes can be understood as follows. The IFE scales as the square of the magnitude of the in-plane type SOC, which is perturbatively incorporated into DSMs resulting in the mass gap of ~10 meV. In contrast, the CGF-like contribution is linear to the SOC and also scales with the interaction energy that characterizes Fermi velocity of the Dirac cone, which is hundreds of meV as derived by the effective continuum model, $\beta = (E^2\ v\Delta_{\mathrm{SO}})/\Omega^3$ (Sec. III). From the discussion here, the dominant contribution of the observed AHC is the CGF-like one arising from the topological property of the DSM, which inevitably leads to the creation of Weyl nodes when a sufficiently strong field is applied.

## V. CONCLUSION

We observed the CPL-induced AHC in the paramagnetic DSM phase of $Co_3Sn_2S_2$ by conducting MIR pump-THz Faraday rotation probe spectroscopy. From the field-strength and driving frequency dependence, the observed AHC is attributed to CPL-induced nonzero Berry curvature associated with the splitting of the Dirac bands as predicted by the Floquet theory. The estimated splitting of the Dirac bands reaches about 60 % of the mass gap and the calculated CPL-induced AHC quantitatively reproduces the experimental observation. Our results demonstrate a promising route toward the realization of Floquet-Weyl states from massive Dirac semimetals with realistic improvements of e.g. the MIR light source and will open further opportunities to investigate dynamically emergent topologically protected properties such as chiral anomaly and surface Fermi arc. The on-demand topological phase transition by light enables ultrafast switching between the 3D Dirac and Weyl state and provides a unique pathway for the ultrafast quantum manipulation of topological materials.

## ACKNOWLEDGEMENT

This work was supported by JST CREST grant no. JPMJCR19T3, JPMJCR18T2, Japan. A.M. acknowledges the support by the US Department of Energy, Office of Science, Basic Energy Sciences under Award No. DE-SC0010821.

N.Y. and S.O. contributed equally to this work.

## APPENDIX A: BAND CALCULATIONS OF THE PARAMAGNETIC PHASE OF $Co_3Sn_2S_2$

We carried out first-principles calculations of the paramagnetic phase of $Co_3Sn_2S_2$, based on the noncollinear density functional theory (DFT) using the quantum-ESPRESSO code [64], and analyzed the band structures using the Wannier-interpolated Hamiltonian [65]. The details of the method and results for the ferromagnetic state of $Co_3Sn_2S_2$ can be found in Ref. [66]. Figure 6(a) represents the calculated band structure of $Co_3Sn_2S_2$ in the paramagnetic phase. Figure 6(b) displays the energy separations of the relevant two bands. The high symmetry points are labeled by magenta circles. The energy separation has a minimum value of ~10 meV at $(k_x, k_z) = (0.44, 0.03)$ Å$^{-1}$ and $(0.73, 0.37)$ Å$^{-1}$, and these massive DPs are indicated by white circles. The line-cuts across the DP are plotted along $k_x, k_y, k_z$ directions in Fig. 6(c)-6(e), respectively. From these dispersions, the group velocities around the DP are estimated to be $\boldsymbol{v} = (v_x, v_y, v_z) = (0.84, 0.85, 1.67)$ eVÅ.

## APPENDIX B: MID-INFRARED PUMP AND THz FARADAY ROTATION MEASUREMENT

The setup for the mid-infrared (MIR) pump-terahertz (THz) Faraday probe measurement is illustrated in Fig. 7. We used a Yb:KGW-based regenerative amplifier (pulse energy 2 mJ, center wavelength 1030 nm, maximum repetition rate 3 kHz, and pulse duration 170 fs in FWHM) as a light source. Half of the output was used to generate the MIR pump pulse. The other half was split into two beams for the probe THz pulse generation and for the electro-optic sampling of the THz pulse, respectively. All the optical measurements were performed at room temperature.

The probe THz pulse was generated by an optical rectification in a 380-μm-thick GaP (110) crystal, and the THz probe pulse transmitted through the sample was detected by using an electro-optic sampling method with another GaP (110) crystal. The Faraday rotation signal was obtained by measuring the THz electric field with the polarization parallel ($E_x$) and perpendicular ($E_y$) to the incident polarization, by changing the configuration of the wire-grid polarizers (WGP1, WGP2 and WGP3 in Fig. 7). The dynamics of the Faraday rotation angle in Fig. 1(d) were obtained by measuring $\Delta\theta = \Delta E_y(t)/E_x(t)$ with the sampling pulse fixed at the peak of the probe THz pulse. The plotted data in Fig. 1(d) for $\sigma^+$ and $\sigma^-$ polarization are obtained by subtracting $\Delta E_y(t)$ for the linearly polarized excitation from those for $\sigma^+$ and $\sigma^-$ polarization to eliminate the leakage signal of the helicity-independent $\Delta E_x(t)$ due to the finite extinction ratio of the used THz polarizers. The complex Faraday rotation angle, $\Theta(\omega) = \theta(\omega) + i\eta(\omega)$, where $\theta(\omega)$ is the Faraday rotation angle and $\eta(\omega)$ is the ellipticity, is determined from the Fourier component of the transmitted probe electric fields as, $\Theta(\omega) = E_y(\omega)/E_x(\omega)$. Plotted in Fig. 2 and 3 are $\Delta\Theta(\omega)$ and $\Delta\sigma_{xy}(\omega)$ for the right-hand circular polarization ($\sigma^+$), which are obtained by taking, e.g., $\{\Delta\sigma_{xy,\sigma^+}(\omega) - \Delta\sigma_{xy,\sigma^-}(\omega)\}/2$, where $\Delta\sigma_{xy,\sigma^\pm}(\omega)$ is AHC induced by $\sigma^\pm$ polarized pump. The delay time between the MIR pump pulse and probe THz pulse with respect to the sampling pulse was controlled by the delay stages (DS1 and DS2).

For the pump, the MIR pulse with nine center frequencies ranging from 50 to 120 THz are used. The spectra of the representative ones were displayed in Fig. 8. The MIR pulse was generated by an optical parametric amplifier (OPA) for 73-120 THz and its combination with a differential frequency generation (DFG) for 50-60 THz. The fluence of the MIR pulse was tuned by using two wire-grid polarizers. The polarization state of the pump pulse was controlled by a liquid-crystal variable retarder for 73-79 THz and a ZnSe photoelastic modulator for 50, 55, 60, 100 and 120 THz. The pump pulse was focused on the sample using a $CaF_2$ lens in a normal incident configuration.

To verify the quantitative comparison of the light-induced AHC for various driving frequencies, the spot size and pulse width were set nearly identical as listed in Table 1. This is necessary due to the non-uniformity of the pump excitation both in space and time. Regarding the spot size, light-induced AHC was measured in a spot of the probe THz pulse which is comparable to that of the MIR pump pulse. Therefore, we set the pump spot size at the sample almost identical for various pump frequencies by optimizing the lens pair for the collimation. The spot size of the probe THz pulse was also measured for different series of the measurements as displayed in Table 1, confirming that the spatial overlap condition between the pump and probe was almost unchanged. The spot sizes of both the pump and probe pulses were measured by knife-edge method. The AHC plotted in Fig. 3(a)-(f) in the main text are obtained by using the values in the higher probe frequency (6.2-8.3 meV in photon energy) that attain better spatial overlap.

The pulse width of the pump can also affect the light-induced AHC quantitatively because it is comparable to the time-resolution of the THz AHC measurement which is ideally limited by the pulse width of the optical sampling pulse, 170 fs. Therefore, we made the pulse width for various frequencies nearly equivalent by tuning the group delay dispersions by inserting dispersing media ($CaF_2$, $Al_2O_3$, Ge, and Si) in the optical path before the sample. The listed pulse width was determined by measuring a cross-correlation between the MIR pump and the optical sampling pulse. From the measured spot size, the pulse width and the pulse energy, the peak electric field of the incident MIR pump pulse was estimated, and then, using the measured optical conductivity (Appendix D) the peak electric field inside the $Co_3Sn_2S_2$ film was calculated.

## APPENDIX C: SAMPLE PREPARATION AND CHARACTERIZATION

The 23 nm-thick, *c*-axis-oriented $Co_3Sn_2S_2$ film were grown on double-side polished $Al_2O_3$ (0001) substrates by radio-frequency magnetron sputtering with a cap of an approximately 50 nm-thick insulating $SiO_2$ layer. Detailed information for the fabrication and characterization of the film sample can be found in Ref. [51]. The ferromagnetic transition temperature of the used film samples is identified at $T_C$~185 K.

## APPENDIX D: EQUILIBRIUM OPTICAL CONDUCTIVITY OF $Co_3Sn_2S_2$ FILM SAMPLE

The complex optical conductivity spectrum of the $Co_3Sn_2S_2$ film is displayed in Fig. 9(a), which was measured by THz time-domain spectroscopy (TDS) and Fourier-transform infrared (FTIR) spectroscopy. For the THz-TDS, that for 2.5-6.0 meV was measured using a 100 fs Ti-Sapphire amplifier-based lab-made THz system with ZnTe (110) crystals, and that for 6-30 meV was measured with a 35 fs Ti-Sapphire amplifier-based lab-made THz system with GaP (110) crystals. The spectrum at the infrared region was obtained with an FTIR microscope, by measuring the transmittance and reflectance of the film sample on substrate. The spectrum in the THz region shows a Drude-like response of the conducting electrons as well as two sharp phonon peaks at 12.4 meV and 19.2 meV. Assuming that other contributions such as interband transition rarely contribute to the spectrum in this low-energy region, the conductivity was fitted by a sum of Drude model and two Lorentz functions (dashed lines), which gives the transport relaxation rate of $\hbar\gamma_{\mathrm{tr}} = 81 \pm 9$ meV.

The optical conductivity spectrum of the present thin-film sample is well consistent with that for the bulk $Co_3Sn_2S_2$ [49,67], including the estimated value of transport relaxation rate. Real-part $\sigma_{xx}$ at the infrared region shows the additional contribution peaked at around 0.7 eV, which can be attributed to the interband transitions. Considering multiple reflections in the film sample on the substrate, we calculated the amplitude of the electric field inside the $Co_3Sn_2S_2$ film (Fig. 9(b)) to correctly estimate the driving field strength. Plotted in Fig. 9(c) is a ratio of the absorbed energy of the film sample over the incident energy as a function of photon energy, which is used for the energy injection to the sample by the MIR pump.

## APPENDIX E: NUMERICAL SIMULATIONS OF THE MEASURED NONEQUILIBRIUM OPTICAL CONDUCTIVITY SPECTRA

In the present study, the pump-induced dynamics of the longitudinal and Hall conductivity includes the temporal region which is even faster than the duration of the probe THz pulse and the period of the used THz wave. In this section, we examine how the nonequilibrium optical conductivity can be reasonably extracted in such an ultrafast temporal region. Firstly, we note that the nonequilibrium optical conductivity spectra, $\Delta\sigma_{xx}(\omega)$ and $\Delta\sigma_{xy}(\omega)$, presented in this paper were obtained by the probe scan method as illustrated in Fig. 10(a). The probe THz pulse transmitted through the sample was measured with fixing the delay between the pump MIR pulse and the electro-optic sampling pulse, $t_{\mathrm{pp}}$, and sweeping the probe THz pulse [68–70]. With this method, even if the duration and/or period of the probe pulse is longer than the pump pulse duration, each point on the temporal waveform of the probe THz pulse is measured as it passes through the sample with the same delay time relative to the pump ($t_{\mathrm{pp}}$). To examine how correctly the measured transient optical conductivity spectrum represents the system response, here we consider the measured conductivity $\sigma^m(\omega)$ and its relationship with the actual conductivity $\sigma(\omega)$. Assuming a free-standing thin-film sample, the THz waveform transmitted through the sample, $E_x^t(t)$, with the conductivity $\sigma_{xx}(t)$ and thickness $d$ can be calculated from the following equations:

$$E_x^t(t) = E_x^i(t) - \frac{1}{2} Z_0 d j_x(t) \qquad \text{(E1)}$$

$$j_x(t) = \int_{-\infty}^{t} dt' \sigma_{xx}(t, t') E_x^t(t') \qquad \text{(E2)}$$

where $E_x^i(t)$, $j_x(t)$, $Z_0$ are the incident probe THz waveform, the induced current density and the vacuum impedance, respectively. In the Drude model, the time-domain expression of the conductivity is given by $\sigma(t, t') = K e^{-\gamma(t-t')} \Theta(t - t')$, where $K$ is the total spectral weight and $\gamma$ is the transport scattering rate. $\Theta(t)$ is a step function. To simulate $\sigma_{xx}^m(\omega)$ at $t_{\mathrm{pp}}$, the pump-induced change in the parameter $\gamma$ is introduced as $\gamma(t, t_{\mathrm{pp}})$ as represented in Fig. 10(b). The transmitted THz waveform at the fixed $t_{\mathrm{pp}}$, $E_x^t(t, t_{\mathrm{pp}})$, is calculated by (E1) and (E2). Figure 10(c) represents the integrand of Eq. (E2) with fixed $t'$ as a function of another time $t_R$. To calculate $E_x^t(t, t_{\mathrm{pp}})$, the value at $t$ marked as the filled circle in Fig. 10(c) is involved in the integration in Eq. (E2). We can see that the value of the integrand is different from that with the time-independent $\gamma_{\mathrm{noneq.}}$ (open square in Fig. 10(c)), which causes a discrepancy of the measured nonequilibrium conductivity at $t_{\mathrm{pp}}$, $\sigma_{xx}^m(\omega, t_{\mathrm{pp}})$, from the actual conductivity, $\sigma_{xx}(\omega, t_{\mathrm{pp}})$. However, this discrepancy will be insignificant when the transport relaxation time $\tau = \gamma^{-1}$ is much shorter than the timescale of the pump-induced change of the parameters because the integration of Eq. (E2) is not much contributed by the parameters ($\gamma$ and $K$) at $t - t' \gg \tau$. Similarly, this discrepancy will be insignificant when the duration of the sampling pulse, which in principle limits the time-resolution of the nonequilibrium conductivity measurement, is longer than $\tau$.

Then, we numerically simulated the measured nonequilibrium conductivity spectra $\sigma_{xx}^m(\omega, t_{\mathrm{pp}})$ and compared them with the actual conductivity spectra $\sigma_{xx}(\omega, t_{\mathrm{pp}})$. Figure 11(a) and (b) present the incident probe THz waveform and its power spectrum for the simulation. Considering the discussion in the previous paragraph, we conducted the simulation with two sets of Drude parameters ($K$ and $\gamma$) in equilibrium; one is the parameters for the used $Co_3Sn_2S_2$ film with $\hbar\gamma = 81$ meV, and the other is the model case with the smaller scattering rate, $\hbar\gamma = 4.1$ meV. The equilibrium longitudinal optical conductivity spectra with these parameter sets are displayed in Fig. 11(c). The pump-induced change was assumed to be 3 % increase of $\gamma$ by the pump pulse excitation with its rise time of 0.1 ps as shown in Fig. 11(d). The duration of the sampling pulse was neglected here, so the time-resolution is not limited in this regard in the simulation. The film thickness $d$ is set as 23 nm. The circles in Fig. 11(e) indicate the calculated real part of $\sigma_{xx}^{m}(\omega)$ as a function of $t_{\mathrm{pp}}$ with the parameter set for $Co_3Sn_2S_2$ film. Although there is a slight difference at around $t_{\mathrm{pp}} = 0$ ps, $\sigma_{xx}^{m}(\omega)$ reasonably well reproduces $\sigma(\omega)$ indicated by solid lines. This is reasonable because the transport scattering time $\tau = 8$ fs in this case is much shorter than the rise time of the pump-induced change. On the other hand, $\sigma_{xx}^{m}(\omega)$ with $\hbar\gamma = 4.1$ meV ($\tau = 160$ fs) show a significant discrepancy with the $\sigma(\omega)$ at $0 \lesssim t_{\mathrm{pp}} \lesssim 0.5$ ps as plotted in Fig. 11(f), which indicates that the discrepancy depends on the scattering time as expected from the consideration above.

To simulate the measurement of the anomalous Hall conductivity, Eq. (E1) and (E2) were extended to the following forms which include the $y$ components of the transmitted THz waveform, $E_y^t(t)$, and the current density, $j_y(t)$, perpendicular to the incident polarization:

$$\begin{cases} E_x^t(t) = E_x^i(t) - \frac{1}{2} Z_0 d j_x(t) \\ E_y^t(t) = -\frac{1}{2} Z_0 d j_y(t) \end{cases} \tag{E3}$$

$$\begin{cases} j_x(t) = \int_{-\infty}^{t} dt' \sigma_{xx}(t,t') E_x^t(t') - \int_{-\infty}^{t} dt' \sigma_{xy}(t,t') E_y^t(t') \\ j_y(t) = \int_{-\infty}^{t} dt' \sigma_{xx}(t,t') E_y^t(t') + \int_{-\infty}^{t} dt' \sigma_{xy}(t,t') E_x^t(t') \end{cases} \tag{E4}$$

The time-domain expression of the intrinsic anomalous Hall conductivity is modeled as $\sigma_{xy}(t,t') = \delta(t-t')\sigma_{xy}^{0}$, where $\delta(t)$ denotes the delta function and $\sigma_{xy}^{0}$ is the dc anomalous Hall conductivity. This instantaneous response provides the frequency-independent value of the $\mathrm{Re}\,\sigma_{xy}(\omega)$ in equilibrium. The measured optical Hall conductivity spectra, $\sigma_{xy}^{m}(\omega, t_{\mathrm{pp}})$, were calculated assuming that the time-dependence of the anomalous Hall conductivity follows the Gaussian profile of the pump pulse. Figure 11(g) displays the $\mathrm{Re}\,\sigma_{xy}^{m}(\omega, t_{\mathrm{pp}})$ at $\hbar\omega = 8.3$ meV as a function of $t_{\mathrm{pp}}$, which shows a good agreement with the dynamics of the actual optical Hall conductivity, $\mathrm{Re}\,\sigma_{xy}(\omega, t_{\mathrm{pp}})$. The frequency-independent nature at various delay times is also confirmed in Fig. 11(h). Although the measured dynamics should be stretched by the finite duration of the sampling pulse (170 fs) which is not incorporated in the present analysis, the simulations confirm that the measured anomalous Hall conductivity reflects well the dynamics of interest even shorter than the duration and period of the probe THz pulse, making it possible to argue the results based on the measured optical conductivity spectra.

## APPENDIX F: THE MIR PUMP INDUCED CHANGE IN THE LONGITUDINAL OPTICAL CONDUCTIVITY

Figure 12(a) displays the dynamics of the MIR-pump-induced change in the transmitted probe THz electric field ($\Delta E_x$) divided by its equilibrium value ($E_x$) for the left-hand circular polarization ($\sigma^-$, red) and right-hand circular polarization ($\sigma^+$, blue). The photon energy and fluence of the MIR pump pulse were set to 0.31 eV and 2.4 $\mathrm{mJ\,cm^{-2}}$, respectively, which are the same as those for the data in Fig. 1(d) in the main text. The signal shows the maximum value at 0.6 ps, followed by a decaying component with a lifetime of more than 5 ps. The grey line indicates the difference between $\Delta E_x/E_x$ for the $\sigma^-$ excitation and that for the $\sigma^+$ excitation, which suggests the absence of the helicity-dependent component in the dynamics of $\Delta E_x/E_x$.

The longitudinal optical conductivity spectra, $\sigma_{xx}(\omega)$, with various pump fluences are displayed for those under the MIR pulse irradiation ($t_{\mathrm{pp}} = 0$ ps, Fig. 12(b) and (c)) and those for the long-lived state ($t_{\mathrm{pp}} = 3$ ps, Fig. 12(d) and (e)). The dashed lines present their equilibrium values. The conductivity spectra at $t_{\mathrm{pp}} = 3$ ps show the up to 3 % decrease of $\mathrm{Re}\,\sigma_{xx}$. This spectral weight missing in the low frequency range can be caused by a decrease in the carrier density or an increase of the carrier scattering rate. Since it is unlikely that the MIR pump picked off the free carriers at the Fermi surfaces and the created nonequilibrium (not quasi-thermal) state lives for 3 ps, the decrease of $\mathrm{Re}\,\sigma_{xx}$ is presumably attributed to the increase of the carrier scattering rate. The spectra at $t_{\mathrm{pp}} = 0$ ps shows a similar change as those at $t_{\mathrm{pp}} = 3$ ps.

Figure 13(a) and (b) show the fluence dependence of $\mathrm{Re}\,\Delta\sigma_{xx}$ with various driving frequencies at $t_{\mathrm{pp}} = 0$ ps and 3 ps, respectively. Note that the plotted data are averaged values of $\mathrm{Re}\,\Delta\sigma_{xx}$ by $\sigma^-$ excitation and by $\sigma^+$ excitation, while $\mathrm{Re}\,\Delta\sigma_{xy}$ plotted in the main text are half of differences of $\mathrm{Re}\,\Delta\sigma_{xy}$ by $\sigma^-$ excitation and by $\sigma^+$ excitation. By line fittings for the weak fluence data, the driving frequency dependence of the pump-induced change in the longitudinal conductivity was displayed in Fig. 13(c)-(e). In Fig. 13(c), the data is plotted normalized by the peak electric field strength, which indicates the frequency dependence is quite different from that of $\mathrm{Re}\,\Delta\sigma_{xy}$ plotted in Fig. 3(g). Since the $\mathrm{Re}\,\Delta\sigma_{xx}$ can be attributed to the energy injection to the sample by a linear light absorption, the data is replotted normalized by the absorbed energy in Fig. 11(d), showing nearly frequency-independent nature. Here the absorbed energy of the film sample was estimated by using the plot in Fig. 9(c) obtained by the infrared optical conductivity.

Plotted in Fig. 13(e) is same as Fig. 13(d) but for $t_{\mathrm{pp}} = 3$ ps, which also confirms a nearly frequency-independent nature.

## APPENDIX G: CALCULATION OF THE ANOMALOUS HALL CONDUCTIVITY IN THE FLOQUET STATE

Based on the effective continuum model introduced in the main text, we calculated the anomalous Hall conductivity (AHC) spectrum in the Floquet state. Here, the effect of light propagation direction, which is tilted from the separation of the pair of DPs in the experiment as illustrated in Fig. 14(a), is taken into account by applying rotational operation around the $y$-axis with an angle $\phi$. The modified Hamiltonian for the massless DSM state is written as,

$$H_{\mathrm{eq}}(\mathbf{k}) = [m - Bk'^2]\tau_z s_0 + v k'_x \tau_x s_0 + \Delta_{\mathrm{SO}} k'_y \tau_y s_z$$
$$= [m - Bk^2]\tau_z s_0 + v(k_x \cos\phi + k_z \sin\phi)\tau_x s_0 + \Delta_{\mathrm{SO}} k_y \tau_y s_z\,, \quad \text{(G1)}$$

where $(k'_x, k'_y, k'_z) = (k_x \cos\phi - k_z \sin\phi\,, k_y, k_z \cos\phi + k_x \sin\phi)$.

By the same procedure as that for Eq. (2), the time-independent effective Hamiltonian under the circularly polarized light (CPL) driving field is given as,

$$H_{\mathrm{eff}}(\mathbf{k}) \approx \left[(\widetilde{M} - Bk^2)s_0 - \beta\cos\phi\, s_z\right]\tau_z + v_y k_y \tau_y + (v_x k_x - v_z k_z)\tau_x \quad \text{(G2)}$$

with $\widetilde{M} = m - \frac{BE^2}{\Omega^2}, \beta = \frac{E^2 v \Delta_{\mathrm{SO}}}{\Omega^3}, v_y = \Delta_{\mathrm{SO}} s_z + \frac{2E^2}{\Omega^3} Bv\cos\phi\, s_0, v_x = v\cos\phi\, s_0 + \frac{2E^2}{\Omega^3} B\Delta_{\mathrm{SO}} s_z$ , and $v_z = v\sin\phi\, s_0$. Then the two DPs split into the four Weyl points (WPs) at

$$\mathbf{K}_{s,\pm} = \left( \pm\sqrt{\frac{\widetilde{M}_s v_z^2}{B(v_z^2 + v_{x,s}^2)}}, 0, \pm\sqrt{\frac{\widetilde{M}_s v_{x,s}^2}{B(v_z^2 + v_{x,s}^2)}} \right) \quad \text{(G3)}$$

Where $s = \pm 1$ represents the $s_z$ component and $\widetilde{M}_s = \widetilde{M} - s\beta\cos\phi$, $v_{x,s} = v\cos\phi + s\frac{2E^2}{\Omega^3} B\Delta_{\mathrm{SO}}$, and $v_z = v\sin\phi$. We can check the chirality of the WPs by expanding the effective Hamiltonian in Eq. (G2) in the vicinity of $\mathbf{K}_{s,\pm}$ in Eq. (G3) as $H^{s\pm}{}_{\mathrm{eff}}(\mathbf{k}) \sim \mathbf{v}^{s\pm} \cdot (\mathbf{k} - \mathbf{K}_{s,\pm})$ with $\mathbf{v}^{s\pm} = (v_x^{s\pm}\sigma_x, v_y^{s\pm}\sigma_y, v_z^{s\pm}\sigma_z)$. The chirality $\chi$ is obtained by the sign of the product of each component as $\chi = \mathrm{sgn}(v_x^{s\pm} v_y^{s\pm} v_z^{s\pm})$, where $\mathrm{sgn}(v_x^{s\pm}) = +1$, $\mathrm{sgn}(v_y^{s\pm}) = s$, and $\mathrm{sgn}(v_z^{s\pm}) = \mp 1$ for $v > 0$ and $\Delta_{\mathrm{SO}} > 0$. The WP with $\chi = +1(-1)$ corresponds to the source (sink) of the Berry curvature as shown in Fig. 14 (b).

Next, the in-plane type of SOC, $H_{\mathrm{SO},x} = m\tau_y s_x$, is added to $H_{\mathrm{eq}}$ in eq. (G1) to introduce the mass gap. By referring the band structure calculation of $Co_3Sn_2S_2$, we adopted a material parameter set, $M = 0.13$ eV, $v = 0.82$ eV Å, $\Delta_{\mathrm{SO}} = 0.85$ eV Å, $B = 1.44$ eV Å$^2$, $\phi = 40°$, and $m = 0.005$ eV. The optical Hall conductivity is calculated from linear response theory [24,53], which leads to

$$\sigma_{\lambda\nu}(\omega) = -\epsilon^{\lambda\nu\rho}\frac{e^2}{\hbar a}\sum_{\alpha\neq\beta}\int\frac{d\mathbf{k}}{(2\pi)^3}\frac{\left(E_{\alpha\mathbf{k}}-E_{\beta\mathbf{k}}\right)^2}{(\hbar\omega)^2-\left(E_{\alpha\mathbf{k}}-E_{\beta\mathbf{k}}\right)^2}f_\alpha(\mathbf{k})[\nabla_{\mathbf{k}} \times \boldsymbol{\mathcal{A}}_\alpha(\mathbf{k})]_\rho \quad \text{(G4)}$$

where $\lambda, \nu, \rho = x, y, z$, $\epsilon^{\lambda\nu\rho}$ is the Levi-Civita symbol, $e$ is the elementary charge, $\hbar$ is the reduced Plank constant, $a = 5.37$ Å is the lattice constant, and $\alpha, \beta$ represent band indices. $f_\alpha(\mathbf{k}) = (1 + e^{(E_{\alpha\mathbf{k}}-\mu)/T})^{-1}$ is the Fermi distribution function and $\boldsymbol{\mathcal{A}}_\alpha(\mathbf{k})$ represents the Berry connection. We introduce the angular frequency of the probing light $\omega$ to compute the ac-response. The temperature $T$ is set to 300 K and the chemical potential $\mu$ is set to 0.12 eV below the center of the mass gap. The calculated real and imaginary parts of the $\sigma_{xy}(\omega)$ spectra are displayed in Fig. 4(d) and (e). The calculated spectra show a nearly constant value in the real part and a negligibly small value in the imaginary part, which well matches the experimentally observed spectral feature as shown in Fig. 2(b) and (c) in the main text. Since the real part in this energy range can be approximately regarded as a constant, we present the calculated value of $\sigma_{xy}(\hbar\omega/2\pi = 10\ \mathrm{meV})$ in Fig. 4(f) and (g) in the main text.

## APPENDIX H: EFFECTIVE TIGHT-BINDING MODEL FOR DESCRIBING THE FLOQUET STATE OF $Co_3Sn_2S_2$

We explain the minimal tight-binding model of the DSM state in non-magnetic $Co_3Sn_2S_2$ and the derivation of the effective Floquet Hamiltonian. The tight-binding model describes the three pairs of the gapped Dirac nodes in the non-magnetic state above $T_{\mathrm{C}}$ [55,56]. In this model, the unit cell consists of the (3+1) sublattices as each one of the *d* orbitals from three Co atoms forming the Kagome layers and one *p* orbital from Sn on the interlayer triangular lattice are extracted. We neglect the rest of all the orbitals for simplicity. The effect of those neglected sites is substantially incorporated into the model as an electric field yielding an SOC, as explained in the following. Here, lattice parameters are given as $a = 5.37$Å and $c = 13.18$Å [39].

The reduced tight-binding Hamiltonian in the real space reads

$$H_{\rm TB} = H_{\rm hop} + H_{\rm KM} + H_{\rm sR}. \qquad \text{(H1)}$$

The spin-independent $H_{\rm hop}$ is given by

$$H_{\rm hop} = -\sum_{ijs} t_{ij} d_{is}^{\dagger} d_{js} + \sum_{\langle ij \rangle s} t_{ij}^{dp} \left( d_{is}^{\dagger} p_{js} + p_{is}^{\dagger} d_{js} \right) + \sum_{is} \epsilon_p p_{is}^{\dagger} p_{is}, \qquad \text{(H2)}$$

where $d_{is}$ and $p_{is}$ are the annihilation operators of the Co-$d$ and Sn-$p$ orbital electrons, respectively. The index $s$ is for the spin, and $i$ for the site. $t_{ij}$ includes the first- and second-nearest-neighbor hopping, $t_1$ and $t_2$, in the kagome layer, and inter kagome layer hopping $t_z$. The summation for $\langle ij \rangle$ is taken over the nearest-neighbor hopping between the Co and Sn sites. $t_{ij}^{dp}$ represents the $d - p$ hybridization between the Co-$d$ and Sn-$p$ orbitals. The hopping vectors incorporated in this model are described in Ref. [54–56].

The Kane-Mele SOC $H_{\rm KM}$ and the staggered Rashba-type SOC $H_{\rm sR}$ are respectively given by,

$$H_{\rm KM} = -it_{\rm KM} \sum_{\langle\langle ij \rangle\rangle ss'} \nu_{ij} d_{is}^{\dagger} \sigma_{ss'}^{z} d_{js'} \qquad \text{(H3)}$$

and

$$H_{\rm sR} = -it_{\rm sR} \sum_{\langle ij \rangle ss'} \boldsymbol{\lambda}_{ij} \cdot d_{is}^{\dagger} \boldsymbol{\sigma}_{ss'} d_{js'}, \qquad \text{(H4)}$$

where $\boldsymbol{\sigma} = (\sigma^x, \sigma^y, \sigma^z)$ is the vector of Pauli matrices, corresponding to the electron spin. The sign $\nu_{ij}$ takes $+1\,(-1)$ when the electron hops counterclockwise (clockwise) to reach the second-nearest-neighbor site on the kagome plane denoted by $\langle\langle ij \rangle\rangle$. The Kane-Mele SOC originating from the nuclei potential of Sn at the center of hexagons of the kagome lattice preserves the $z$ component of the spin operator and gaps out the nodal ring except for the Dirac points. On the other hand, the staggered Rashba-type SOC $H_{\rm sR}$ originates from the local breaking of inversion symmetry of the Sn and S sites [71]. We use $\boldsymbol{\lambda}_{ij}$ to denote the effective magnetic field vector from the electric field perpendicular to the Kagome plane when the electron hops from the site $j$ to $i$, where $\boldsymbol{\lambda}_{AB} = \left(\frac{1}{2}, \frac{\sqrt{3}}{2}, 0\right), \boldsymbol{\lambda}_{BC} = (-1,0,0), \boldsymbol{\lambda}_{CA} = \left(\frac{1}{2}, -\frac{\sqrt{3}}{2}, 0\right)$. The summation $\langle ij \rangle$ is taken for intralayer nearest-neighbor sites. This staggered Rashba-type SOC plays an important role in obtaining the band gap for the Dirac nodes in the non-magnetic state.

In Fig. 5(a), we show the energy spectrum obtained by momentum representation of the Hamiltonian $H_{\mathrm{TB}}(\mathbf{k})$. Here, $H_{\mathrm{TB}} = \sum_{\mathbf{k}} C_{\mathbf{k}}^{\dagger} H_{\mathrm{TB}}(\mathbf{k}) C_{\mathbf{k}}$, where $C_{\mathbf{k}}$ is the set of the momentum represented annihilation operators with sublattice indices A, B, and C for Co sites, given as $C_{\mathbf{k}} = (d_{A\uparrow\mathbf{k}}, d_{B\uparrow\mathbf{k}}, d_{C\uparrow\mathbf{k}}, p_{\uparrow\mathbf{k}}, d_{A\downarrow\mathbf{k}}, d_{B\downarrow\mathbf{k}}, d_{C\downarrow\mathbf{k}}, p_{\downarrow\mathbf{k}})$.
For our calculations, we set $t_1 = 0.15$ eV, $t_2 = 0.8\, t_1$, $t_z = -1.0\, t_1$, $t_{dp} = 1.8\, t_1$, $\epsilon_p = -7.2\, t_1$, and $t_{\mathrm{KM}} = t_{\mathrm{sR}} = 0.2\, t_1$. These parameters are also chosen to fit the Weyl points structure in ferromagnetic states to the result obtained by the first-principles calculations.

Based on this tight-binding model, we derive the time-independent Floquet Hamiltonian with a similar procedure in the continuum model. We again mention that the effect of circularly polarized laser is incorporated by applying the Peierls substitution in momentum space as $H_{\mathrm{TB}}(t) = H_{\mathrm{TB}}(\mathbf{k} + e\mathbf{A}(t))$. We adopt this process for all the hopping terms in Eq. (H1) except for the on-site potential term. With the high-frequency expansion for the first order of $\Omega$, the effective Floquet Hamiltonian for each crystalline momentum is given by Eq. (2) in the main text, with the $n$th component of the Floquet Hamiltonian. For example, the explicit form for the Floquet Hamiltonian for $t_1$ hopping term is given by,

$$
\begin{aligned}
&H_n(\mathbf{k}) \\
&= -2t_1 J_n(Aa) \begin{pmatrix} 0 & e^{i\frac{2}{3}n\pi}\cos(\mathbf{k}\cdot\mathbf{b}_1) & \cos(\mathbf{k}\cdot\mathbf{b}_2) & 0 \\ e^{-i\frac{2}{3}n\pi}\cos(\mathbf{k}\cdot\mathbf{b}_1) & 0 & e^{-i\frac{2}{3}n\pi}\cos(\mathbf{k}\cdot\mathbf{b}_3) & 0 \\ \cos(\mathbf{k}\cdot\mathbf{b}_2) & e^{i\frac{2}{3}n\pi}\cos(\mathbf{k}\cdot\mathbf{b}_3) & 0 & 0 \\ 0 & 0 & 0 & 0 \end{pmatrix} \otimes I
\end{aligned}
\tag{H5}
$$

and

$$H_n(\mathbf{k}) = -2it_1 J_n(Aa)\begin{pmatrix} 0 & e^{i\frac{2}{3}n\pi}\sin(\mathbf{k}\cdot\mathbf{b}_1) & \sin(\mathbf{k}\cdot\mathbf{b}_2) & 0 \\ e^{-i\frac{2}{3}n\pi}\sin(\mathbf{k}\cdot\mathbf{b}_1) & 0 & e^{-i\frac{2}{3}n\pi}\sin(\mathbf{k}\cdot\mathbf{b}_3) & 0 \\ \sin(\mathbf{k}\cdot\mathbf{b}_2) & e^{i\frac{2}{3}n\pi}\sin(\mathbf{k}\cdot\mathbf{b}_3) & 0 & 0 \\ 0 & 0 & 0 & 0 \end{pmatrix}\otimes I \tag{H6}$$

for even and odd *n*, respectively [72]. Here, $J_n(x)$ is the Bessel function of the first kind with order *n* which satisfies $J_{-n}(x) = (-1)^n J_n(x)$ and *I* is the 2×2 identity matrix for the spin degree of freedom. Other terms can be calculated in a similar manner. In contrast to the continuum model, the second term of Eq. (2) with $n \geq 2$ can be finite. However, in our numerical calculations, we truncate this summation at $n = 5$ because no quantitative difference is found by adding terms of larger $n$. From those frameworks, we calculate the energy band structure of the effective Floquet Hamiltonian in Fig. 5(a) and (b) by using the exact diagonalization.

We also calculate the AHC for the effective Floquet-Hamiltonian for the tight-binding model by using linear response theory as

$$\sigma_{xy}(\omega) = \frac{e^2}{\hbar}\sum_{\alpha\neq\beta}\int_{\mathrm{BZ}}\frac{d\mathbf{k}}{(2\pi)^3}\frac{\left(f_\alpha - f_\beta\right)\left(\varepsilon_\alpha - \varepsilon_\beta\right)}{\hbar\omega + \varepsilon_\alpha - \varepsilon_\beta + i\Gamma}\Omega_{\alpha\beta} \tag{H7}$$

where $\varepsilon_\alpha$ and $|\alpha\rangle$ are $\alpha$ th eigenenergy and eigenstate for the effective Floquet Hamiltonian, $\Gamma$ is the broadening factor reflecting the relaxation time, and $\Omega_{\alpha\beta} = \mathrm{Im}\frac{\langle\alpha|V_x|\beta\rangle\langle\beta|V_y|\alpha\rangle}{\left(\varepsilon_\alpha - \varepsilon_\beta\right)^2}$ with the velocity operator $V_\mu = \frac{\partial H(\mathbf{k})}{\partial k_\mu}$. The distribution function is given by $f_\alpha = \left(1 + e^{(\varepsilon_\alpha - \mu_{\mathrm{eff}})/T}\right)^{-1}$ whose effective chemical potential $\mu_{\mathrm{eff}}$ is determined to conserve the electron filling $N_e$ to be the same as the original band structure for $H_{\mathrm{TB}}$ labeled by $\zeta$: $N_e = \sum_\alpha f_\alpha = \sum_\zeta \left(1 + e^{(\varepsilon_\zeta - \epsilon_{\mathrm{F}})/T}\right)^{-1}$ where $\epsilon_{\mathrm{F}}$ is the Fermi energy in the equilibrium state. In the main text, we set the electron filling $N_e = 3/8$ imitating Ref. [55]. The Berry curvature of the $\alpha$-th band is also given by $\mathcal{B}_\alpha(\mathbf{k}) = \sum_\beta \Omega_{\alpha\beta}(\mathbf{k})$ and Fig. 15(a) shows the distribution in the momentum space in the Floquet-Weyl state with $E = 3.2\ \mathrm{MV\ cm^{-1}}$. We also plot the density of Berry curvature with $E = 1.0\ \mathrm{MV\ cm^{-1}}$ and $3.2\ \mathrm{MV\ cm^{-1}}$ in Figs. 5(c) and 15(b), respectively.

The calculated AHC presented in Fig. 5(e) and (f) shows a good agreement even quantitatively with the experiment, but it shows a factor mismatch. To examine this, we also investigated the model parameter dependence of the CPL-induced AHC in the tight-binding model. Our model parameters are adjusted to reproduce the band structure obtained by first-principles calculations well, but the electron filling may deviate from realistic one since the number of orbitals is limited. Taking this into account, we plot the filling dependence in Fig. 15(c). As this material is a metal, we can see that the AHC, which is the integral value of the Berry curvature, can change significantly depending on the electron filling. In addition, since this first-principles calculation does not consider the change in hopping integral due to electron correlation, it seems that the bandwidth will differ from that of the actual material [73]. The nearest-neighbor hopping dependence is shown in Fig. 15(d). Since the AHC can change by several times due to changes in these model parameters, this is thought to be one of the causes of the factor mismatch between experiment and theory.

## APPENDIX I: FULL FLOQUET CALCULATION OF EFFECTIVE TIGHT BINDING MODEL FOR $Co_3Sn_2S_2$

We investigate the impact of the $\pm n$-photon sectors by considering the "full Floquet Hamiltonian" $H_F$, which is given by

$$H_{\mathrm{F}} = \begin{pmatrix} \ddots & \vdots & \vdots & \vdots & \iddots \\ \cdots & H_0 + \Omega & H_{-1} & H_{-2} & \cdots \\ \cdots & H_1 & H_0 & H_{-1} & \cdots \\ \cdots & H_2 & H_1 & H_0 - \Omega & \cdots \\ \iddots & \vdots & \vdots & \vdots & \ddots \end{pmatrix}. \tag{I1}$$

The Floquet quasi-energy band structure of the tight-binding model under the CPL with $E = 1.0\ \mathrm{MV/cm}$ is displayed in Fig. 16(a). Here, the overlap weight with the original wavefunctions is indicated by the color scale, which is calculated by $w_\alpha = \sum_\zeta |\langle \zeta | \alpha \rangle|^2$ where $|\alpha\rangle$ and $|\zeta\rangle$ are the eigenstates of the Floquet state and $\zeta$-th equilibrium band, respectively. The quasi-energy band structure shows the hybridization of the Floquet bands with different photon numbers, as well as the band splitting within the photon sidebands. We checked that the results are not affected by the truncation of the Floquet matrix for $n > 5$.

For full Floquet Hamiltonian, we also calculate the AHC employing Eq. (H7) by assuming that $f_\alpha$ is represented by the "sudden-approximation" [53], i.e., $f_\alpha = \sum_\zeta |\langle \zeta | \alpha \rangle|^2 \left(1 + e^{(\varepsilon_\zeta - \epsilon_{\mathrm{F}})/T}\right)^{-1}$. Figure 16(b) presents CPL-induced AHC at the probe photon energy of 10 meV as a function of the driving frequency with various broadening factors, $\Gamma$, for full Floquet calculation and high-frequency expansion (HFE). The AHC by the full Floquet calculation shows a nonmonotonic frequency dependence, which is likely attributed to the resonant contributions to the Berry curvature arising from the hybridization of the Floquet bands. On the other hand, the AHC by HFE evaluating the contribution of the Floquet 0-photon bands obeys $\Omega^{-3}$, which agrees with the experimental observation. The frequency-nonmonotonic resonance contribution in the full Floquet calculation is significantly suppressed as increasing $\Gamma$, approaching the AHC from the HFE. In contrast, the calculation by HFE indicates that the AHC originated from the 0-photon state much less depends on $\Gamma$. These trends can also be seen in Fig. 16(c), which plots the $\Gamma$-dependence of AHC for several driving frequencies. Although the relaxation introduced in this calculation is the simplest one within the scope of linear response theory, in reality more complicated relaxation mechanisms such as electron scattering and modulation of distribution functions could also be unfavorable to the resonance effect.

## APPENDIX J: CONSIDERATION OF THE ZEEMAN TERM IN THE FLOQUET STATE

The CPL-induced Zeeman splitting in $Co_3Sn_2S_2$, which is regarded as an inverse Faraday effect (IFE) here, is considered. First, we examined this effect with an effective continuum model developed in the main text (Eq. (2)) to get an intuitive understanding. The Zeeman splitting term that will be proportional to $\tau_0 s_z$ can emerge in the light-induced term $[H_{-1}, H_{+1}]/\Omega$ with high-frequency expansion of Floquet states when the equilibrium Hamiltonian has $\boldsymbol{k}$-dependent $s_x$ and $s_y$ terms. This is because only $\boldsymbol{k}$-dependent terms affect the light-induced term by Peierls substitution, and $s_z$ components are derived from commutators of $s_x$ and $s_y$. Figure 17(a) represents how the band structure is modulated by the effect of CGF and IFE. We here draw a *massless* 3D Dirac state as an initial state for better understanding of the essence. The separation of the emergent WPs is along the $k_z$ momentum direction in the case of CGF, while that is along the energy direction in the case of IFE. In Eq. (2), the in-plane SOC to reproduce the gaps at DPs is chosen as the $k$-independent term ($m\tau_y s_x$), which does not cause a light-induced Zeeman term. However, various descriptions can also be considered as an origin of the mass gap. For example, an additional $\boldsymbol{k}$-dependent in-plane SOC, $H_{\mathrm{SO},xy}(\mathbf{k}) = -\tilde{m}\tau_0(k_x s_x + k_y s_y)$, will cause the Zeeman splitting, the energy of which is estimated as $-\frac{\tilde{m}^2 E^2}{\Omega^3}$ by simply calculating the effective Floquet Hamiltonian of $H_{\mathrm{SO},xy}(\mathbf{k})$ under the CPL irradiation. Here, $\tilde{m} = ma$ where the mass gap energy $m = 5\ \mathrm{meV}$ and the lattice constant $a = 5.37$ Å. The derivation of this IFE is similar to that investigated for Rashba electron systems [61,74]. The calculated Zeeman splitting energy is 0.11 μeV, much smaller than $\beta = 3.1$ meV which is equivalent to the amount of the CGF for the massless 3D DSMs. The AHC induced by this Zeeman term in the continuum model is calculated from Eq. (G4). We confirmed that the induced AHC by this Zeeman term is negligible compared to that by the band splitting due to $\beta$, as shown in Fig. 17(b).

For more realistic evaluation, since the proper description of the $\boldsymbol{k}$-dependent in-plane SOC in the effective continuum model is unclear, we examined the effect of the IFE by the effective tight-binding model for the quantitative evaluation in the observed CPL-induced AHC. Considering that the CPL-induced Zeeman field here originates from the in-plane staggered Rashba type SOC (sR-SOC), $H_{sR}(\boldsymbol{k})$, which is also responsible for the mass gaps, we compared $\sigma_{xy}$ induced in the Floquet states with and without sR-SOC in the equilibrium Hamiltonian. The result in Fig. 17(c) shows that the CPL-induced $\sigma_{xy}$ without sR-SOC is only 7 % smaller than that with it. Note here that the CPL-induced Zeeman field proportional to $s_z$ becomes nonzero only when there exists sR-SOC, but sR-SOC can also affect the band structure modification equivalent to CGF for massless 3D DSMs through, e.g., the modulation of the Fermi surfaces in the equilibrium state. Therefore, 7 % change in $\sigma_{xy}$ possibly includes both IFE and other contributions. However, in any case, the result here indicates that the CPL-induced AHC in the Floquet state of $Co_3Sn_2S_2$ is dominated by the splitting of the Dirac bands with its magnitude equivalent to that of CGF for massless 3D DSMs.

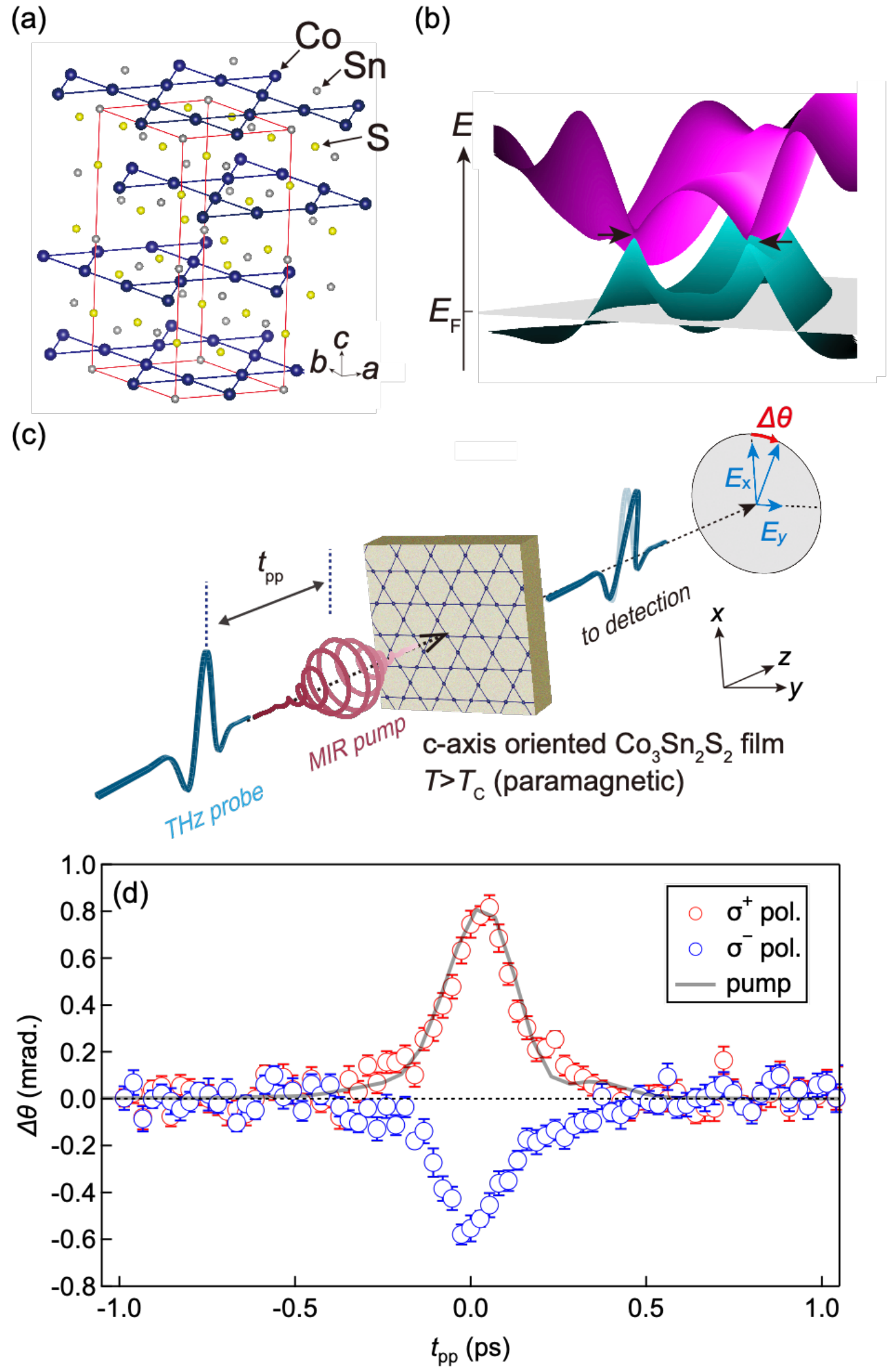

**Figure 1** (a) Crystal structure of $Co_3Sn_2S_2$ where the cobalt atoms form a kagome lattice. (b) Band structure of $Co_3Sn_2S_2$ obtained by the DFT calculation. The massive Dirac points are indicated by black arrows. The grey plane shows the Fermi energy. (c) Schematic of the MIR pump-THz Faraday probe spectroscopy on $Co_3Sn_2S_2$ film. The MIR pump-induced Faraday rotation of the transmitted THz probe pulse, $\Delta\theta$, is measured as a function of pump-probe delay time $t_{\mathrm{pp}}$. (d) Time evolution of MIR pump-induced Faraday rotation angle $\Delta\theta$ of the probe THz pulse transmitted through the $Co_3Sn_2S_2$ film, for right-hand circular polarization ($\sigma^+$, red open circles) and left-hand circular polarization ($\sigma^-$, blue open circles). The grey line indicates the cross-correlation signal of the pump pulse. Error bars indicate the standard errors obtained by multiple measurements.

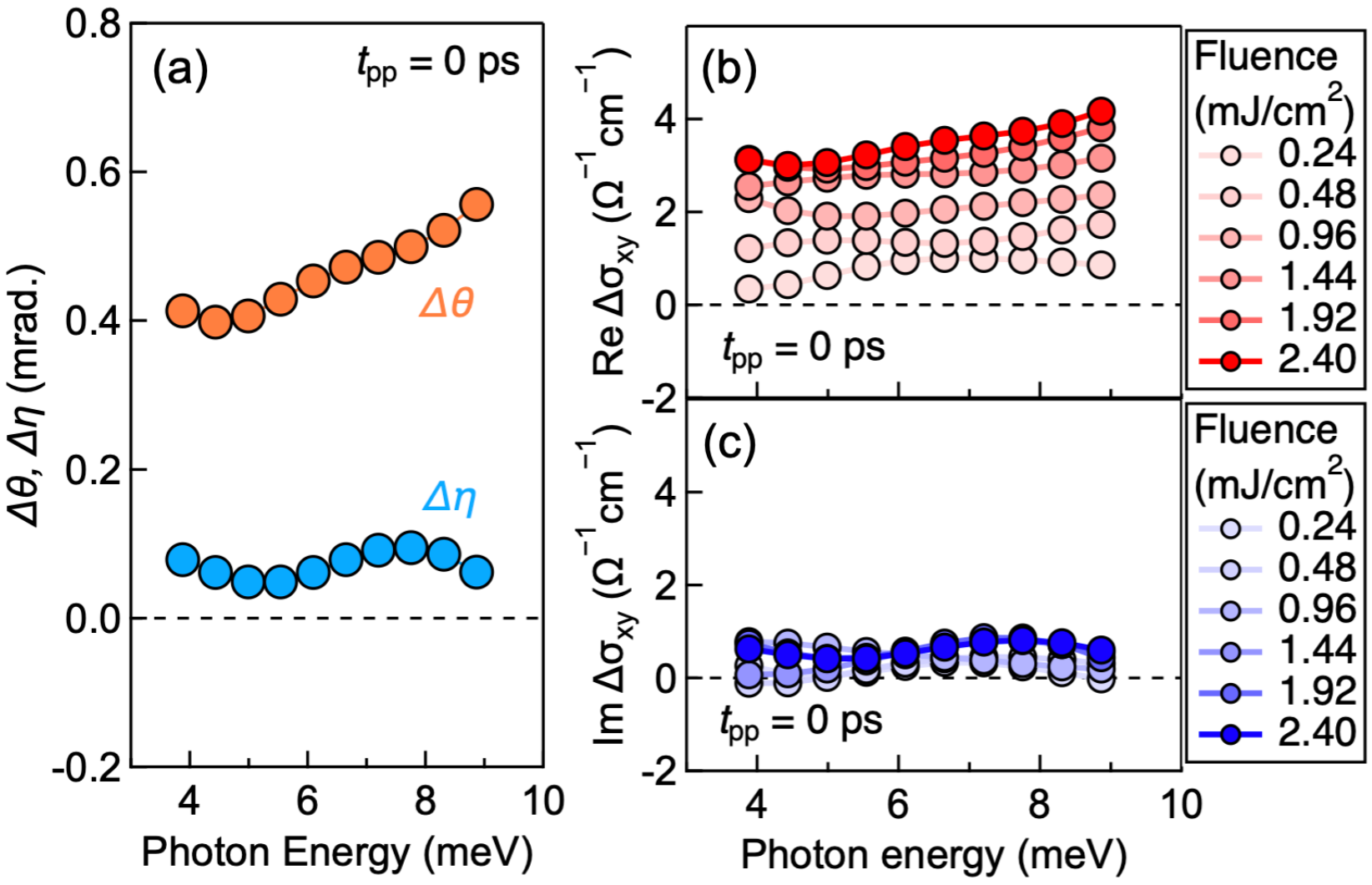

**Figure 2** (a) CPL-induced spectrum of Faraday rotation angle $\theta$ and ellipticity $\eta$ of the THz probe pulse at $t_{\mathrm{pp}} = 0$. (b) Real and (c) imaginary part of the THz AHC spectra, respectively, measured at $t_{\mathrm{pp}} = 0$ with various fluences.

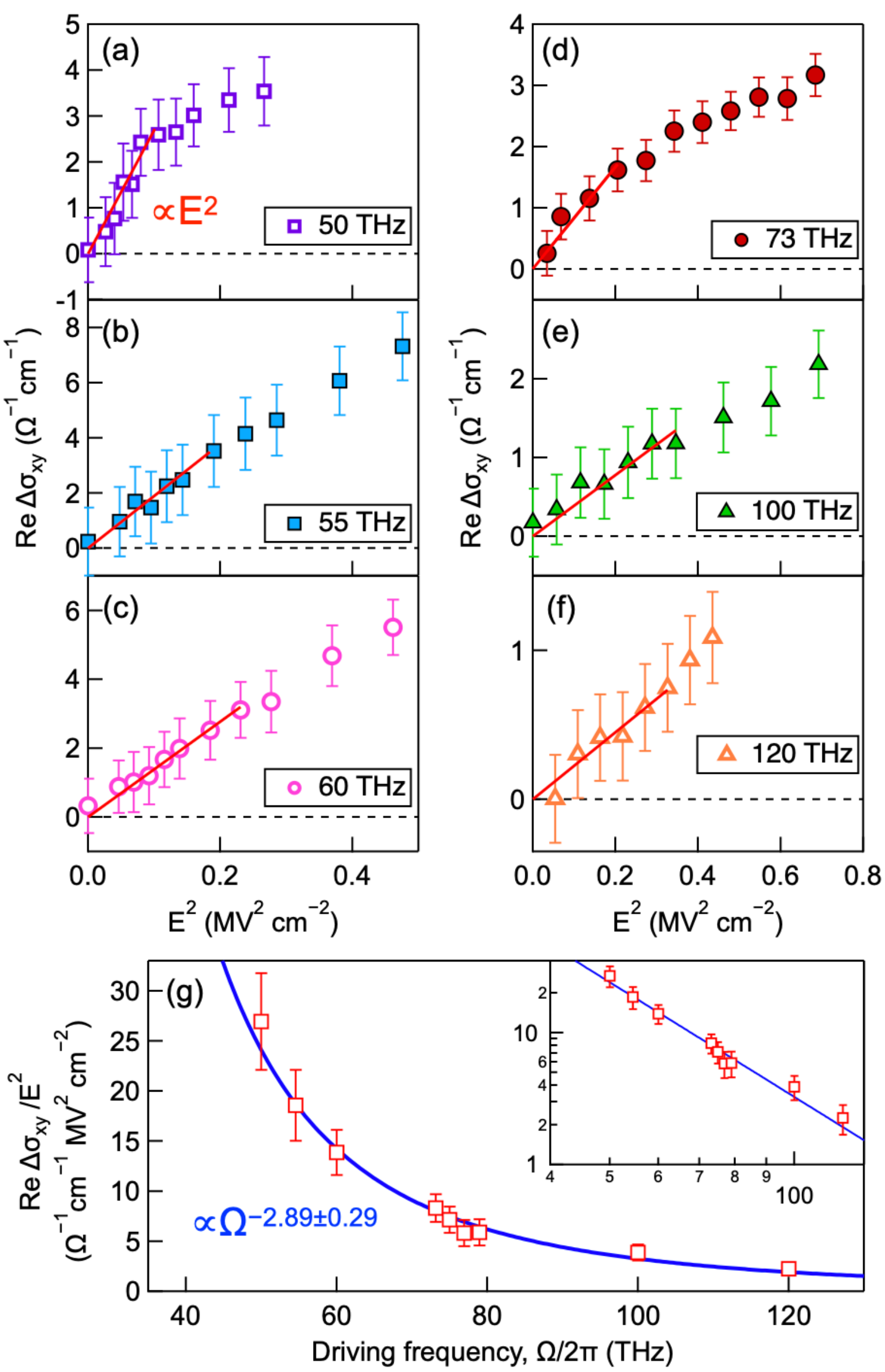

**Figure 3** (a)-(f) Field-strength dependence of the CPL-induced AHC for various driving frequencies. Error bar indicates the standard deviation of multiple measurements. Red lines indicate the line-fitting in the weak field region. (g) Driving frequency dependence of the CPL-induced AHC with the power-law fitting (blue line). The inset shows a log-log plot of the same data.

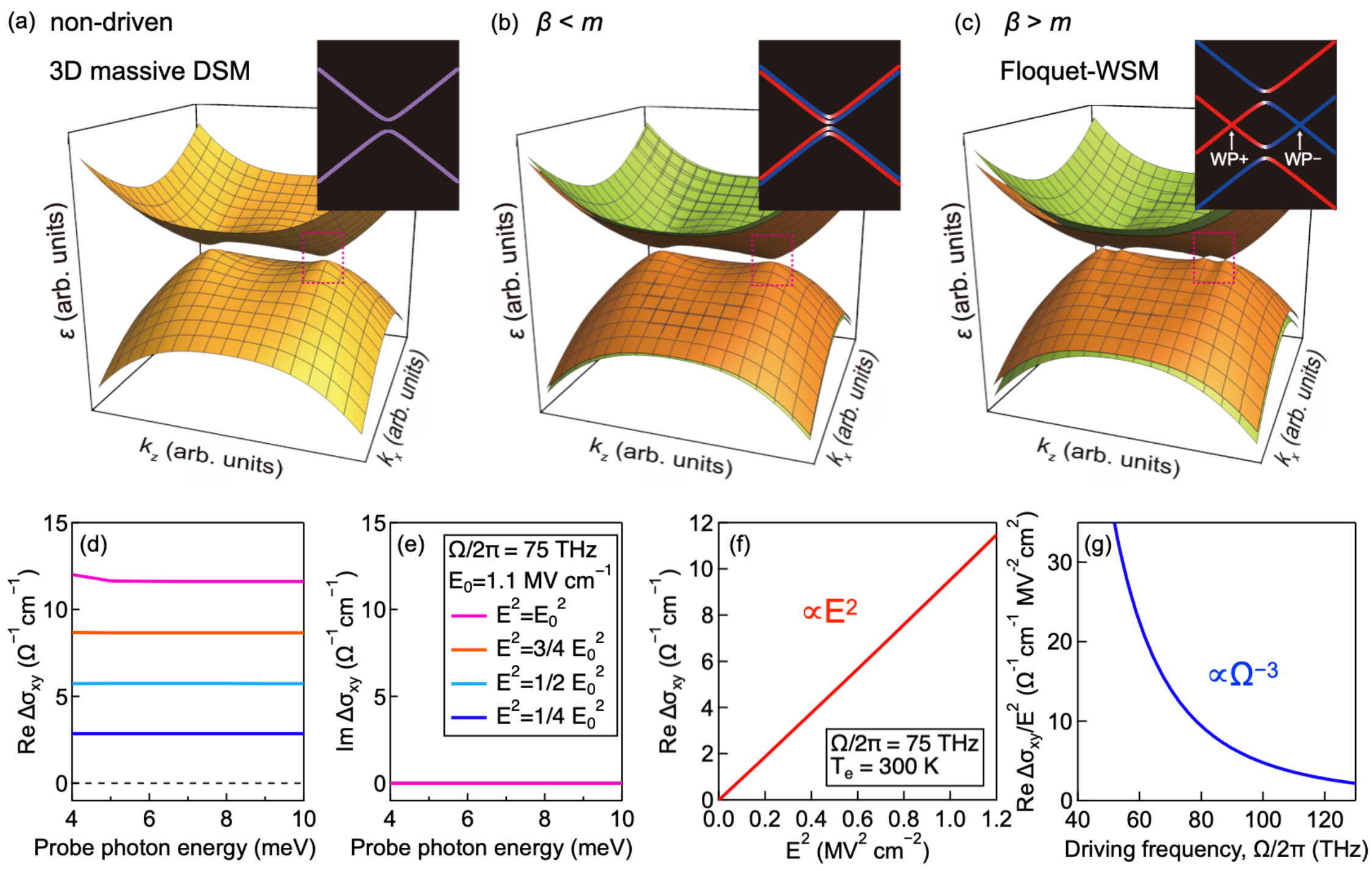

**Figure 4** (a) Band structure derived from the effective continuum model for $Co_3Sn_2S_2$. (b), (c) Band structure of the Floquet state with CPL driving in the case of $\beta < m$ and $\beta > m$. $m$ denotes the mass gap energy. $\beta$ denotes the amount of the CPL-induced band splitting, corresponding to the splitting energy at the massive DPs. The inset in (a)-(c) represents the energy bands around the massive DP as functions of $k_z$. The color of the plotted energy bands reflects the spin state; purple indicates the spin-degenerated band, while red, blue, and white indicate the expectation value of the spin $\langle s_z \rangle = 1, -1$ and 0, respectively. (d) Real and (e) imaginary part of the calculated CPL-induced AHC spectra with various field-strength. (f) Field-strength dependence of the CPL-induced AHC. (g) Driving frequency dependence of the CPL-induced AHC.

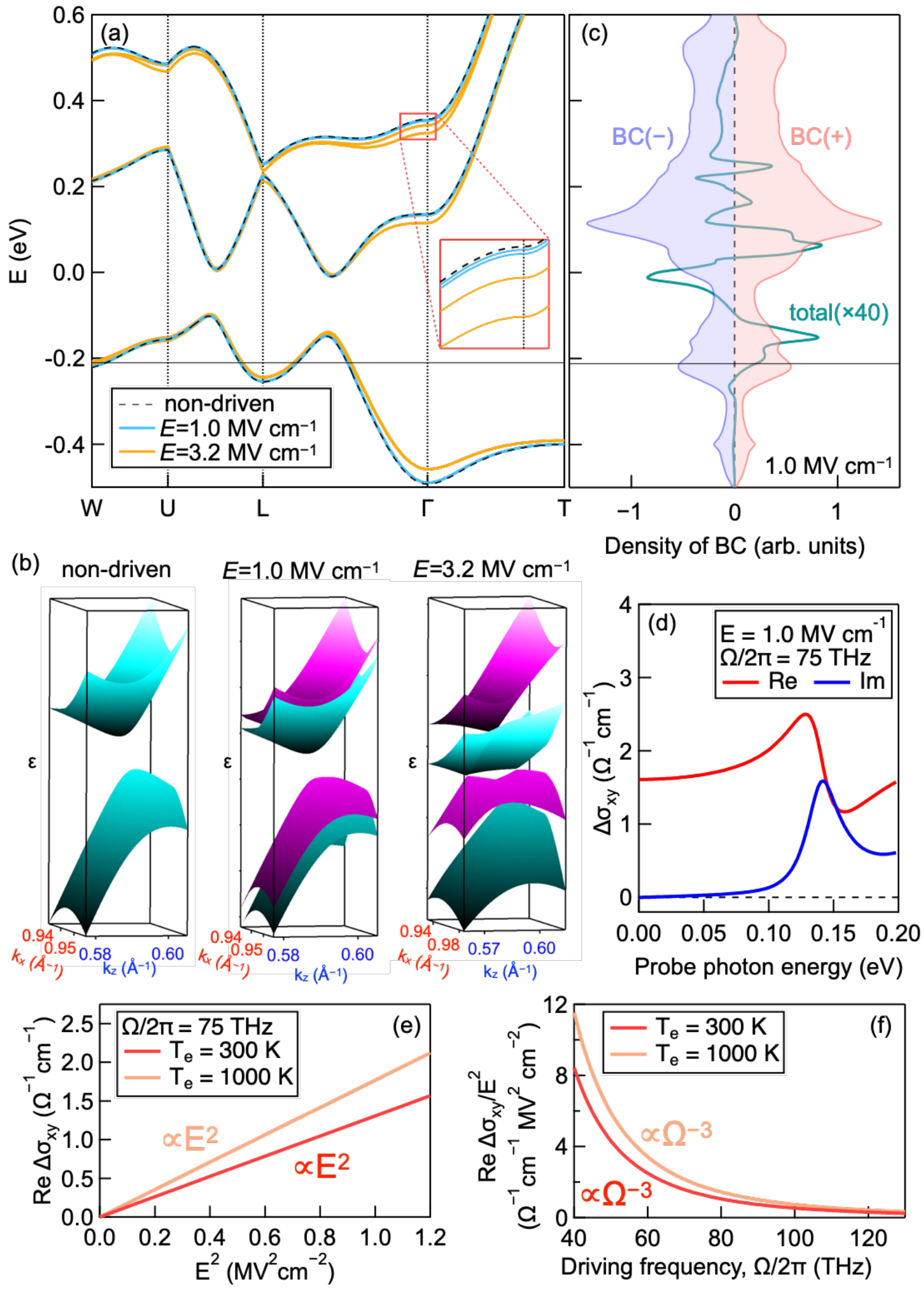

**Figure 5** (a) Band structure of $Co_3Sn_2S_2$ calculated from the tight-binding model in equilibrium (black dashed lines) and with CPL driving for 1.0 (turquoise lines) and 3.2 $\mathrm{MV\,cm^{-1}}$ (orange lines). The driving frequency is set as 75 THz. The inset shows the enlarged view. (b) Enlarged band structures around the DP under the driving electric field of 0, 1.0, and 3.2 $\mathrm{MV\,cm^{-1}}$, respectively. (c) Density of Berry curvature (BC) with the field strength of 1.0 $\mathrm{MV\,cm^{-1}}$. The red (blue) line with the shaded area indicates the positive (negative) component of the BC. The sum of the positive and negative BC is plotted by green line. (d) Calculated AHC spectrum of the Floquet state with the field strength of 1.0 $\mathrm{MV\,cm^{-1}}$. (e) Calculated field-strength dependence and (f) Driving frequency dependence of the CPL-induced AHC with the electron temperature of 300 K (blue) and 1000 K (red).

## Figures in Appendix

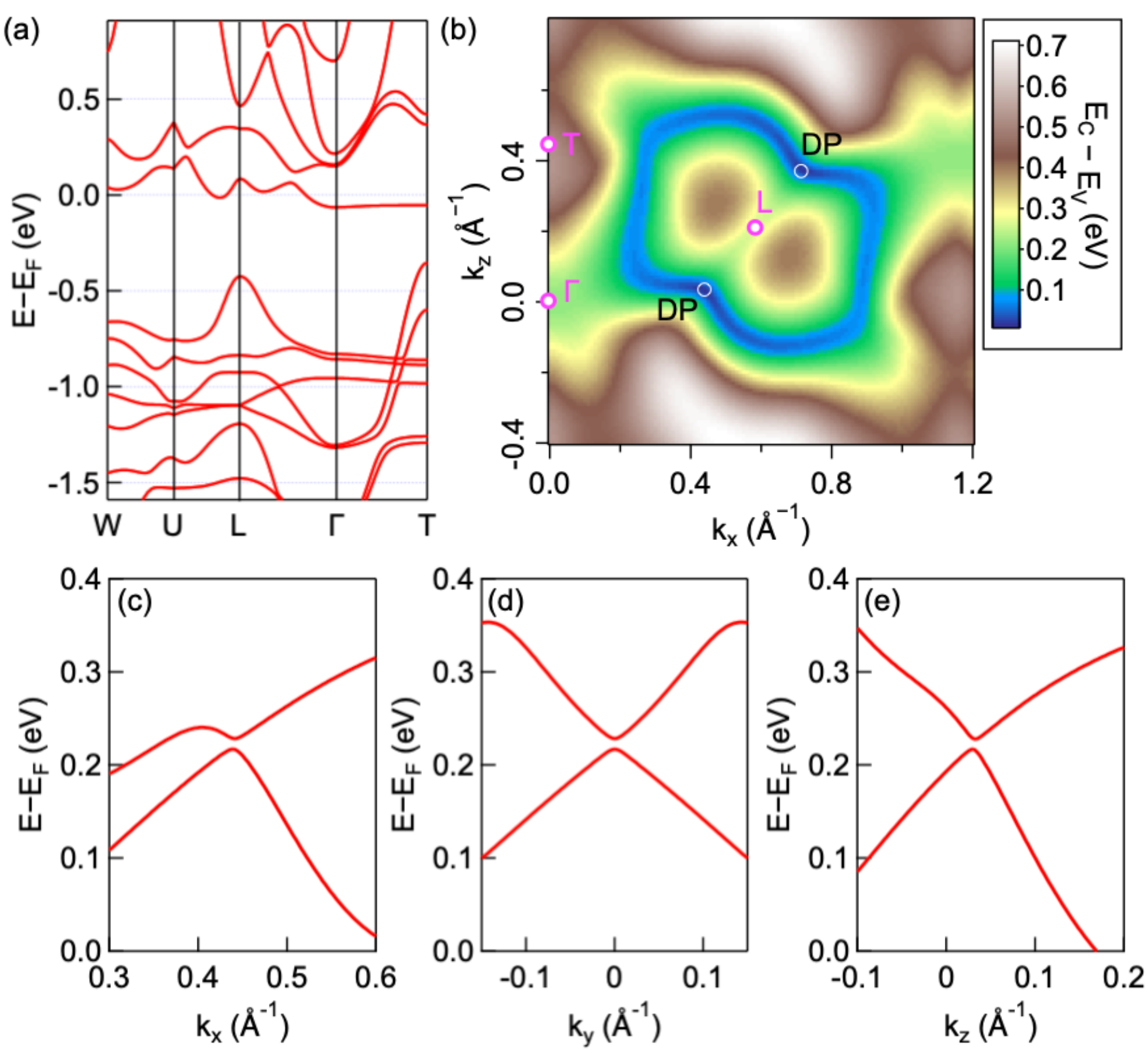

**Figure 6** (a) Band structure of $Co_3Sn_2S_2$ in the paramagnetic phase along high-symmetry points, which was obtained by first-principles calculations. (b) Energy difference of the relevant two bands of $Co_3Sn_2S_2$ in its paramagnetic phase. The high-symmetry points and the Dirac points (DPs) were indicated by circles. (c)-(e) Band dispersions around the massive DP along $k_x, k_y, k_z$ directions, respectively.

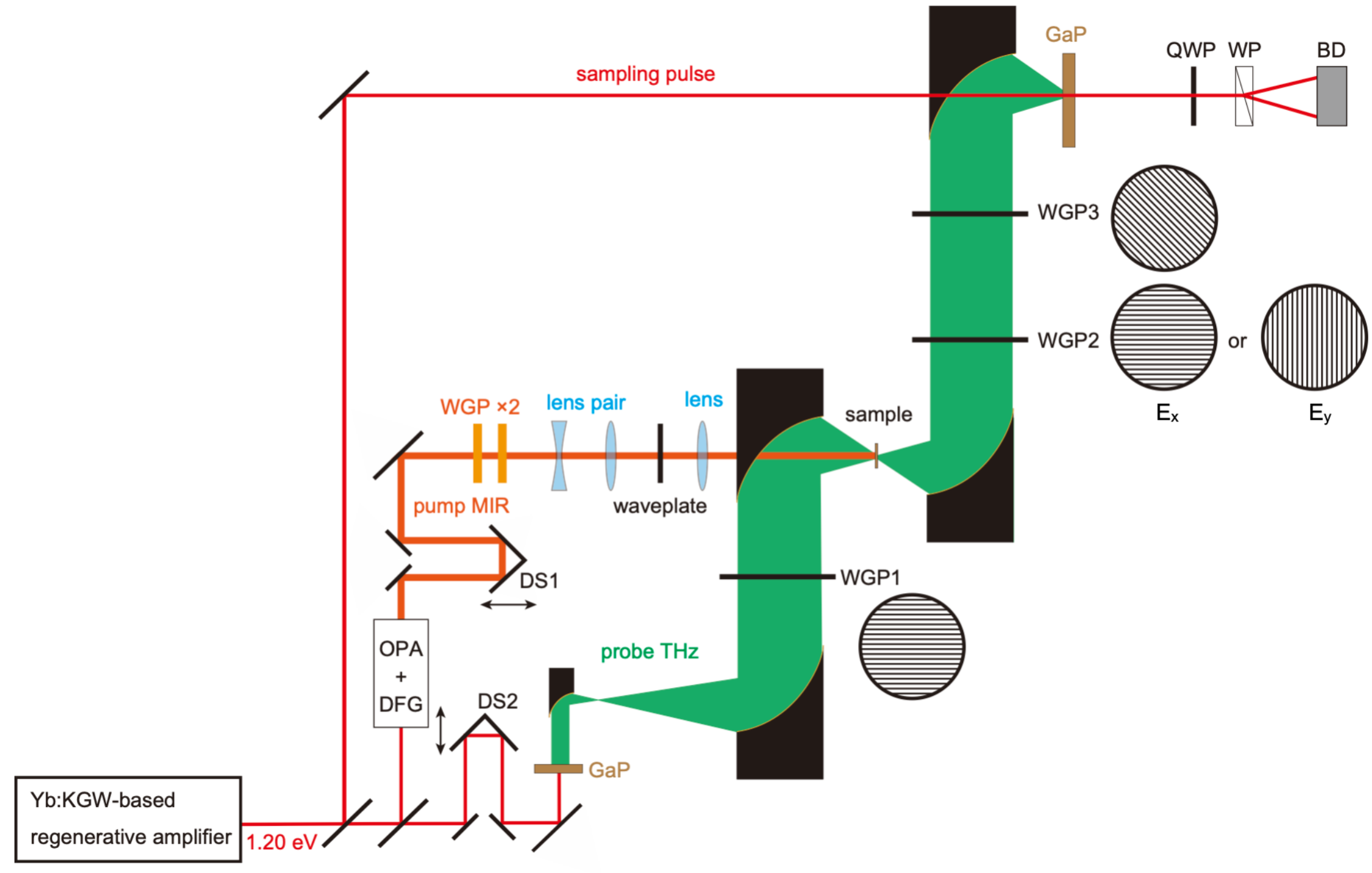

**Figure 7** Experimental setup for the MIR pump-THz Faraday probe measurement. OPA: optical parametric amplifier, DFG: difference frequency generation, DS: delay stage, WGP: wire-grid polarizer, QWP: quarter waveplate, WP: Wollaston prism, BD: balanced photodetector.

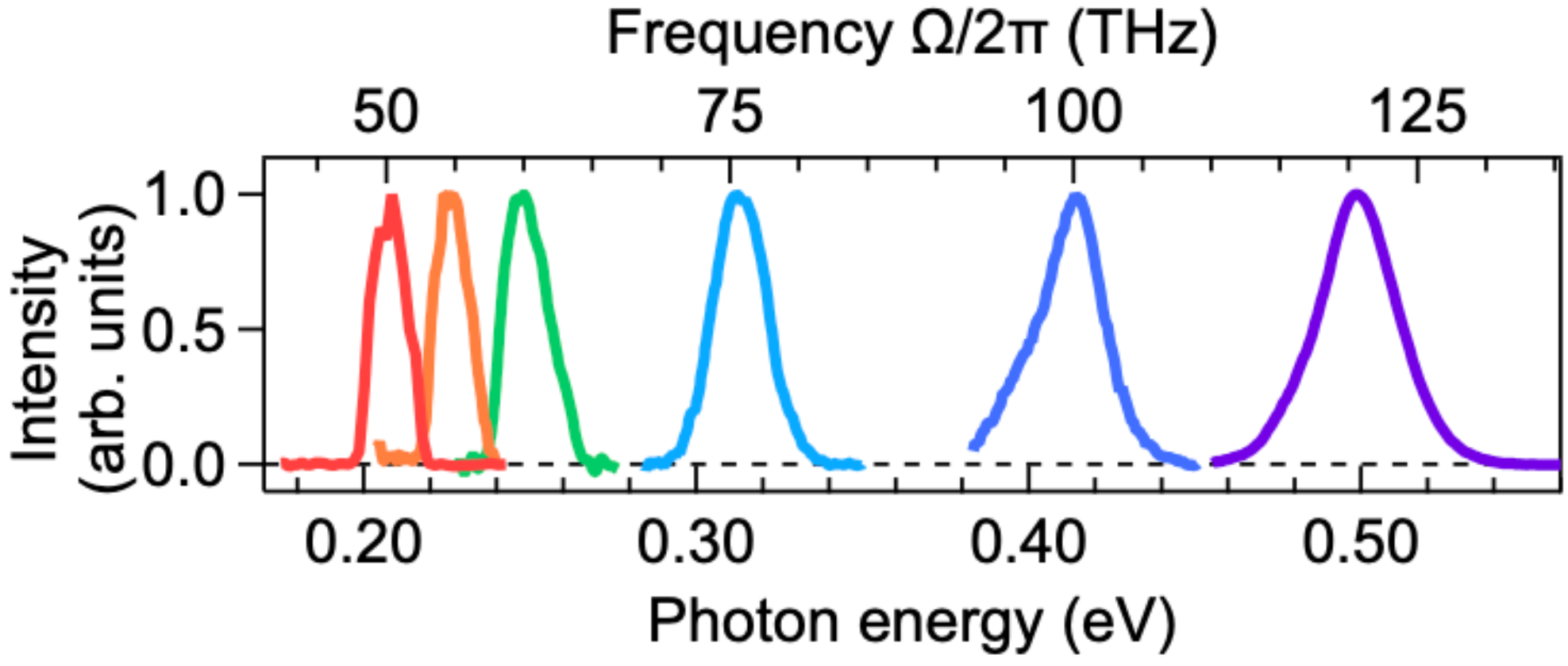

**Figure 8** Spectra of the mid-infrared (MIR) pulses with various center frequencies used as pump pulses.

| Frequency (THz) | Pulse width (fs, FWHM) | Pump spot size (1/e power radius) H(μm)×V(μm) = A(μm²) | Probe spot size at 1 THz (1/e power radius) H(μm)×V(μm) = A(μm²) | Probe spot size at 2THz (1/e power radius) H(μm)×V(μm) = A(μm²) |
|---|---|---|---|---|
| **50** | 189 | 190×194 = 36860 | 270×250=67500 | 180×160=28800 |
| **55** | 181 | 181×198 = 35838 | “ | “ |
| **60** | 169 | 183×196 = 35868 | “ | “ |
| **73** | 172 | 188×189 = 35532 | 240×310=74400 | 200×140=28000 |
| **75** | 160 | 190×185 = 35150 | “ | “ |
| **77** | 190 | 185×189 = 34965 | “ | “ |
| **79** | 216 | 187×182 = 34035 | “ | “ |
| **100** | 162 | 192×183 = 35136 | 290×310=89900 | 170×170=28900 |
| **120** | 160 | 194×186 = 36084 | 290×270=78300 | 170×160=27200 |

**Table 1** Characterization of the pump and probe pulses for the various driving frequencies.

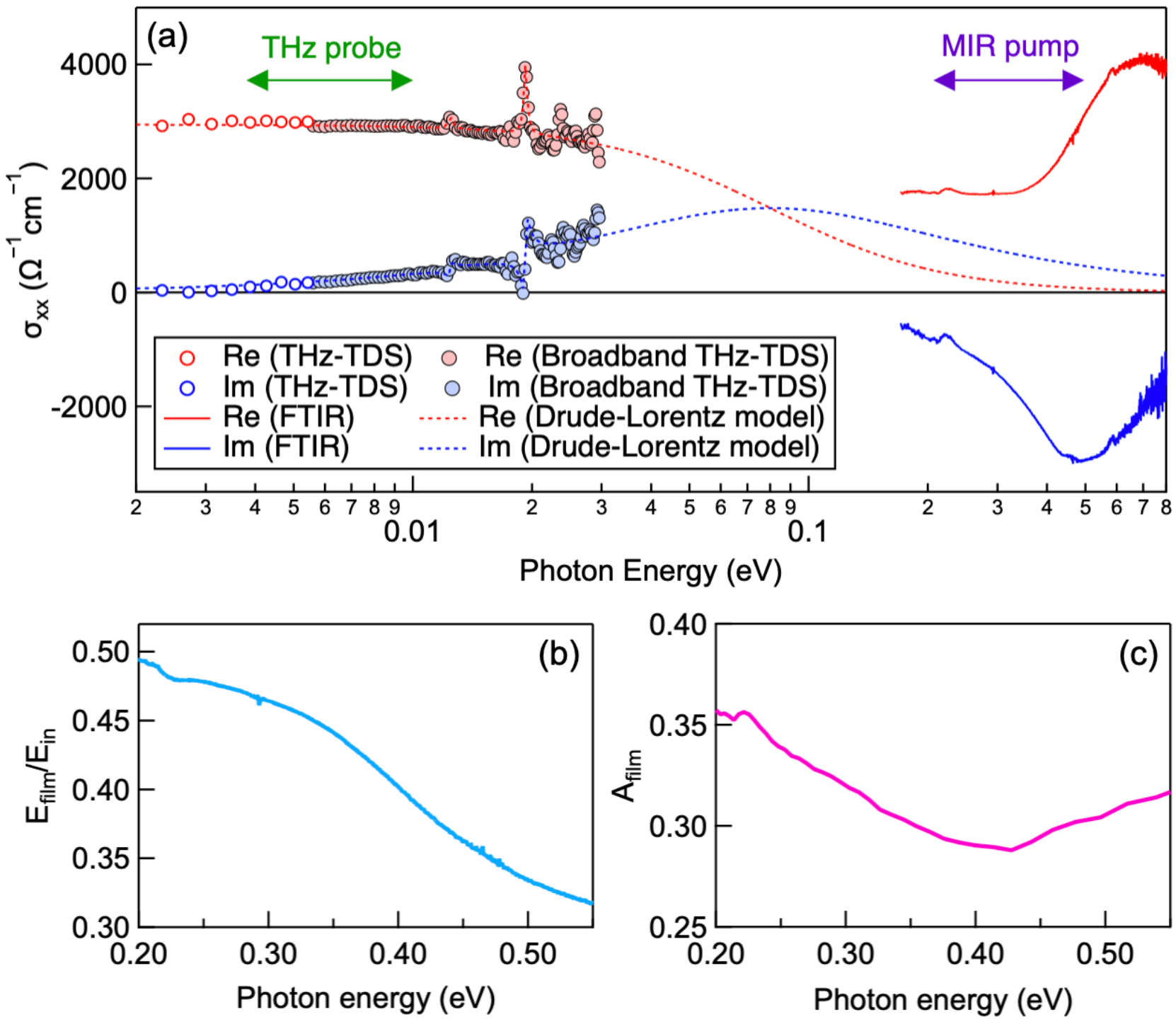

**Figure 9** (a) Conductivity spectrum of the $Co_3Sn_2S_2$ thin film at 300 K, which was measured by the combination of the THz time-domain spectroscopy (THz-TDS), broadband THz-TDS, and Fourier transform infrared (FTIR) spectroscopy. The photon energy ranges of the MIR pump and THz probe are indicated by purple and green arrows, respectively. (b) The electric field amplitude in the $Co_3Sn_2S_2$ film divided by that of the incident light calculated by the conductivity spectrum in (a). (c) Absorption ratio of the $Co_3Sn_2S_2$ film that is the energy absorbed by the film sample divided by the energy of the incident light, calculated by the conductivity spectrum in (a).

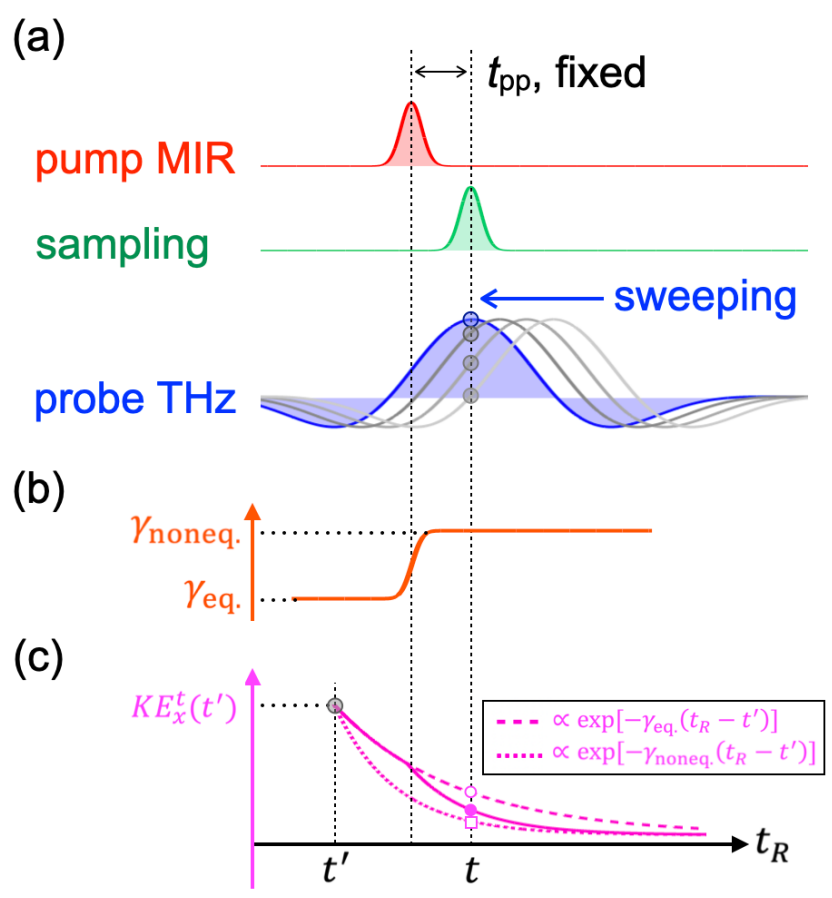

**Figure 10** (a) Schematic of the measurement scheme in the present study. (b) The MIR pump-induced change in the transport scattering rate. (c) The integrand of Eq. (E2) with the constant $\gamma_{eq.}$ (dashed line), the constant $\gamma_{noneq.}$ (dotted line) and time-varying $\gamma$ shown as (b) (solid line).

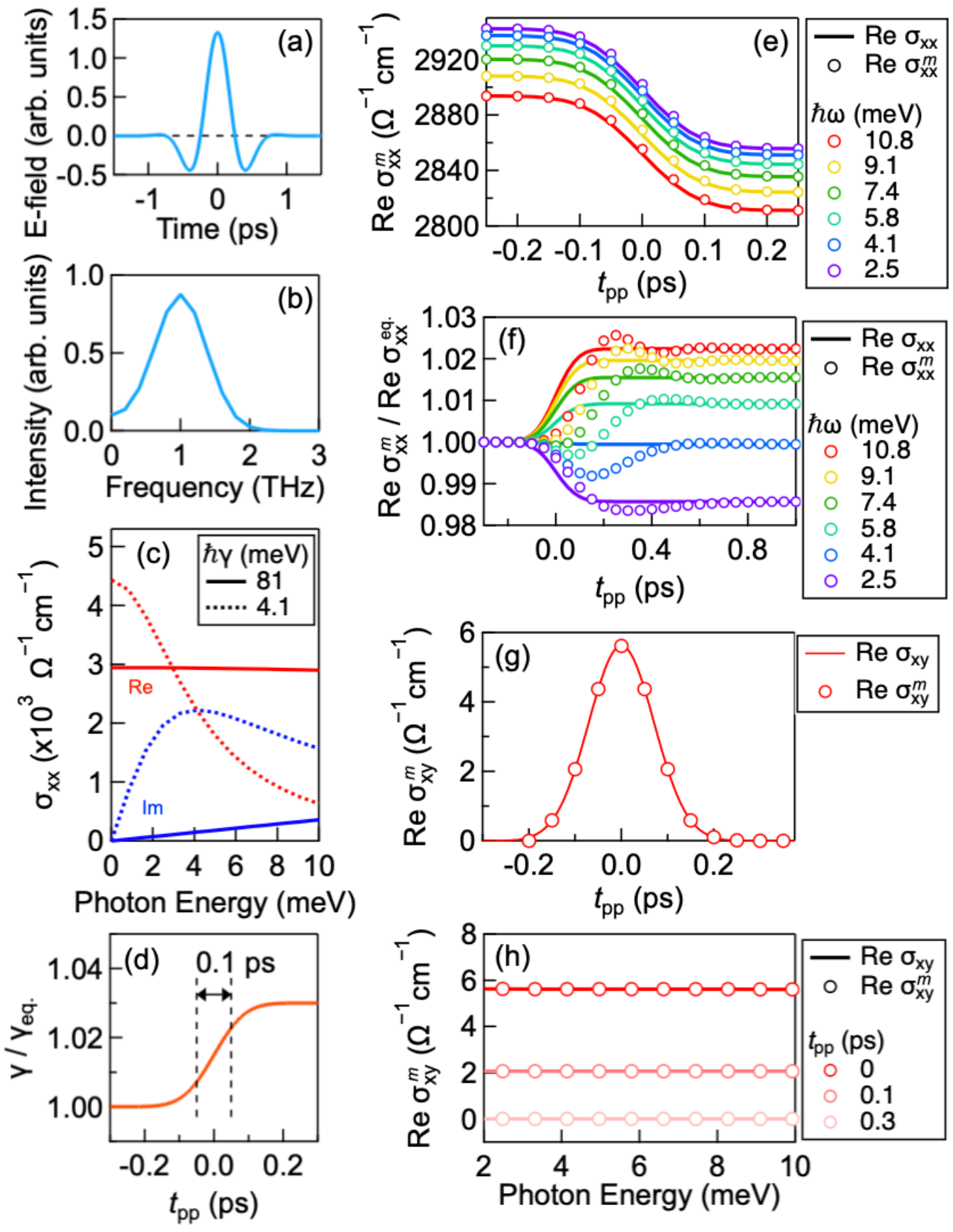

**Figure 11** (a) The waveform and (b) power spectrum of the incident probe THz pulse used in the simulations. (c) Real-part (red) and imaginary-part (blue) of the equilibrium optical conductivity spectra with $\hbar\gamma = 81$ meV (solid line) and $\hbar\gamma = 4.1$ meV (dotted line). (d) The pump-induced change in the transport relaxation rate assumed in the simulations. (e) The real-part of the measured optical conductivity $\sigma_{xx}^{m}$ (circles) and the actual optical conductivity $\sigma_{xx}$ (solid lines) with $\hbar\gamma = 81$ meV, as a function of the pump-probe delay time. The colors indicate the probe photon energy. (f) Same as (e) with $\hbar\gamma = 4.1$ meV. The plotted values are normalized by their equilibrium values for clarity. (g) The real-part of the measured anomalous Hall conductivity $\sigma_{xy}^{m}$ (circles) and the actual optical conductivity $\sigma_{xy}$ (solid line) as a function of the pump-probe delay time. The probe photon energy is 8.3 meV. (h) The spectra of the real-part of the measured anomalous Hall conductivity $\sigma_{xy}^{m}$ (circles) and the actual optical conductivity $\sigma_{xy}$ (solid line) with different pump-probe delay times.

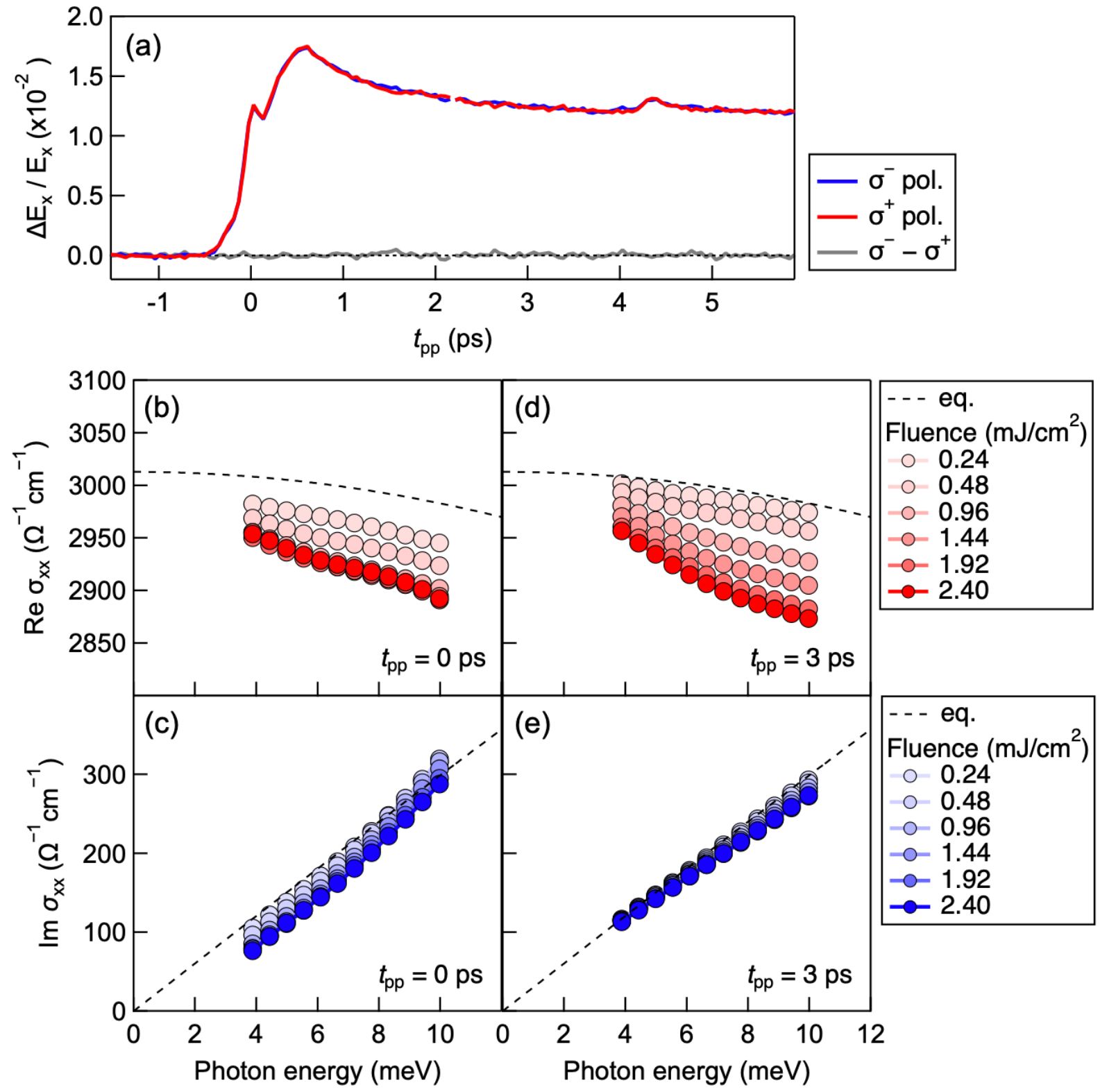

**Figure 12** (a) Time evolution of the transmitted THz electric field $\Delta E_y$ divided by its equilibrium value $E_y$, as a function of the pump-probe delay time $t_{\mathrm{pp}}$. The data for $\sigma^{-}$ (blue), $\sigma^{+}$ (red) excitation, and the helicity dependent component (grey) are plotted. The hump at $t_{\mathrm{pp}}\sim 4$ ps is attributed to the internal reflection of the pump pulse inside the substrate. (b) Real and (c) imaginary part of the pump induced change of the longitudinal optical conductivity $\sigma_{xx}$ at $t_{\mathrm{pp}} = 0$ ps, with the various pump fluences. The equilibrium values are indicated by dashed lines. (d), (e) Same as (b) and (c), but for $t_{\mathrm{pp}} = 3$ ps.

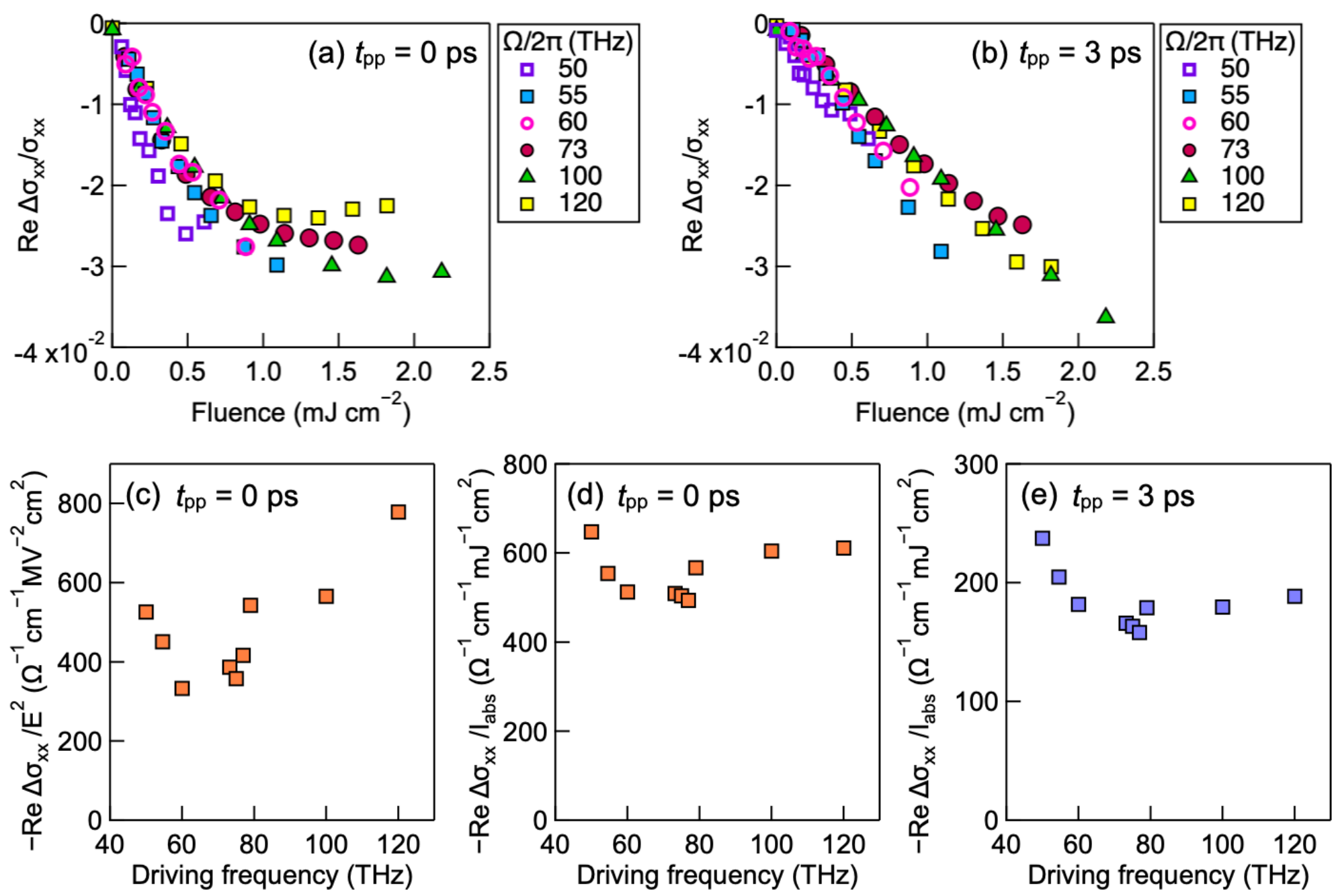

**Figure 13** (a), (b) Fluence dependence of the pump-induced change of the longitudinal optical conductivity at $t_{\mathrm{pp}} = 0$ ps and 3 ps, respectively. (c) Driving frequency dependence of the the pump-induced change of the longitudinal optical conductivity at $t_{\mathrm{pp}} = 0$ ps with a normalization by the squared electric field strength inside the $Co_3Sn_2S_2$ film. (d) Same as (c) but with a normalization by the absorbed energy by $Co_3Sn_2S_2$ film. (e) Same as (d) but for $t_{\mathrm{pp}} = 3$ ps.

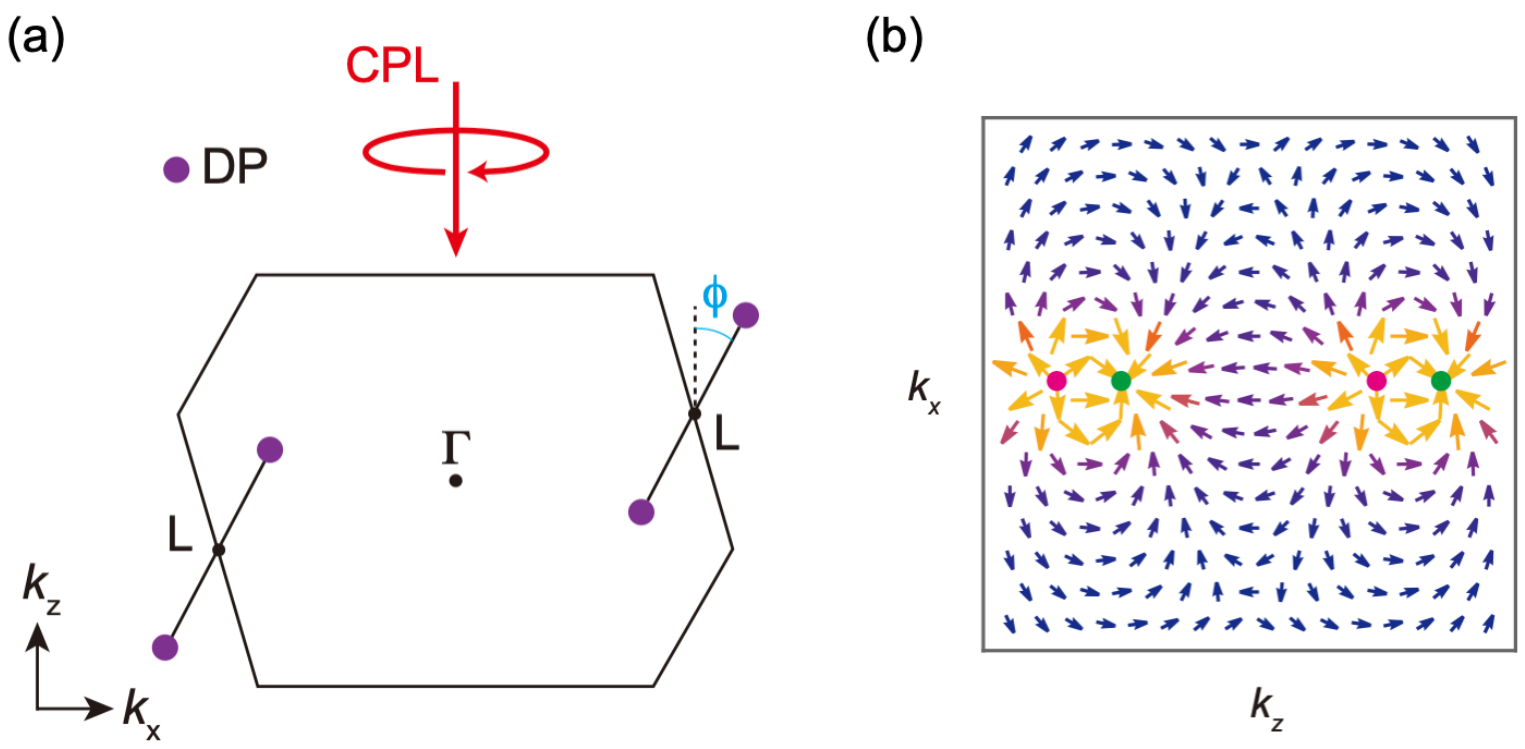

**Figure 14** (a) Illustration of the configuration of the DPs in $Co_3Sn_2S_2$ and the CPL. Black solid hexagon indicates the first Brillouin zone in the $k_y = 0$ plane. The CPL propagates along the c-axis of the crystal while the separation of the DPs is tilted by angle $\phi$. (b) Schematic of the Berry curvature distribution of the CPL-induced Floquet-Weyl state. The WPs with chirality plus (minus) are indicated by magenta (green) circles.

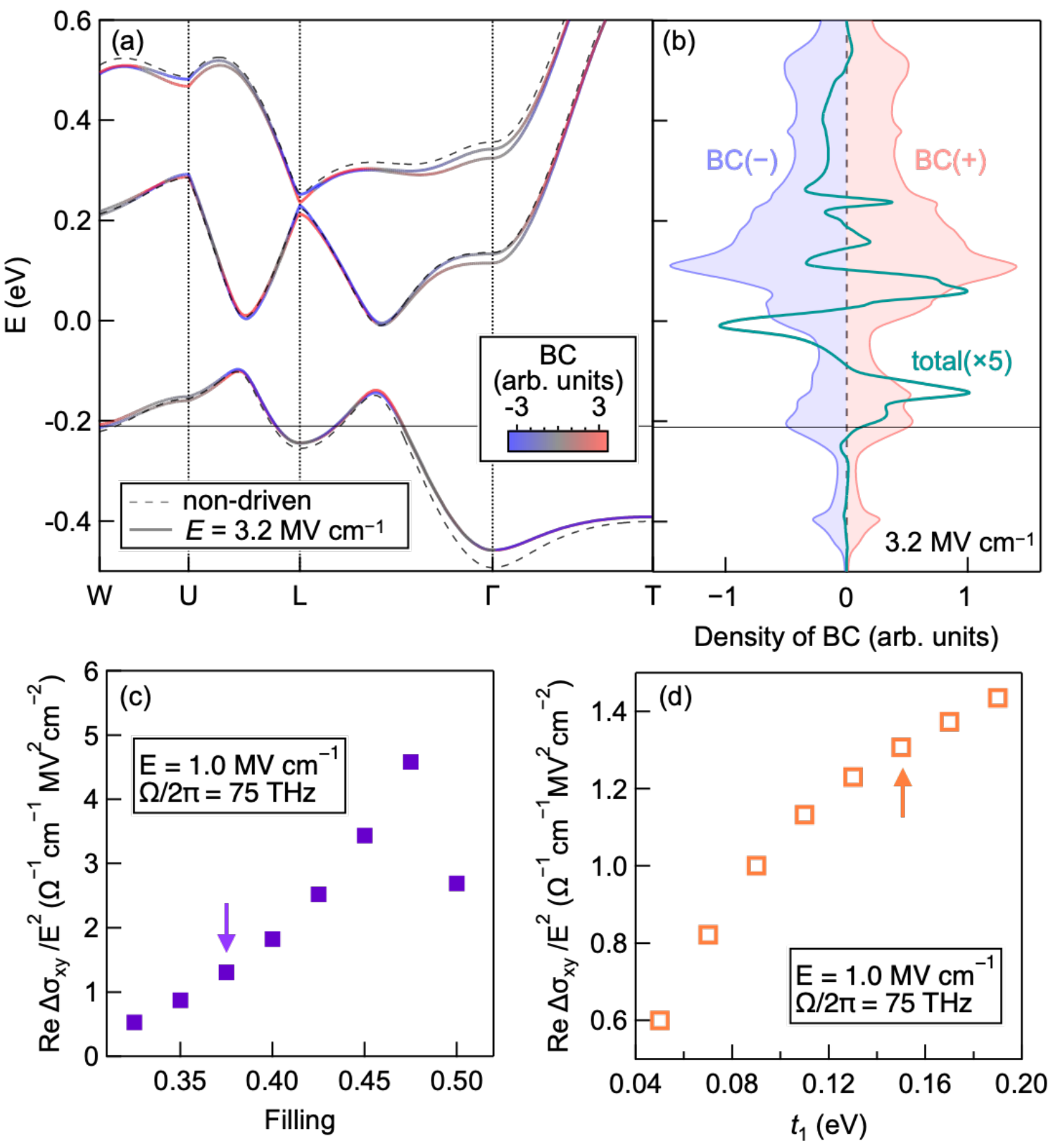

**Figure 15** (a) Floquet band structure of $Co_3Sn_2S_2$ tight-binding model calculated by high-frequency expansion, where the Berry curvature (BC) is indicated by the color scale. The electric field strength and frequency of the driving CPL are set to 3.2 MV/cm and 75 THz, respectively. Dashed lines indicate the non-driven band structure. (b) Density of BC with the field strength of 3.2 MV cm$^{-1}$. The red (blue) line with the shaded area indicates the positive (negative) component of the BC. The sum of the positive and negative BC is plotted by green line. (c) Filling dependence of the CPL-induced AHC. The arrow indicates 3/8 filling which is adapted for the calculations displayed in the main text. (d) Hopping parameter ($t_1$) dependence of the CPL-induced AHC. The arrow indicates $t_1 = 0.15$ eV which is adapted for the calculations displayed in the main text. For (c) and (d), the electric field strength and frequency of the driving CPL are set to 1.0 MV/cm and 75 THz, respectively.

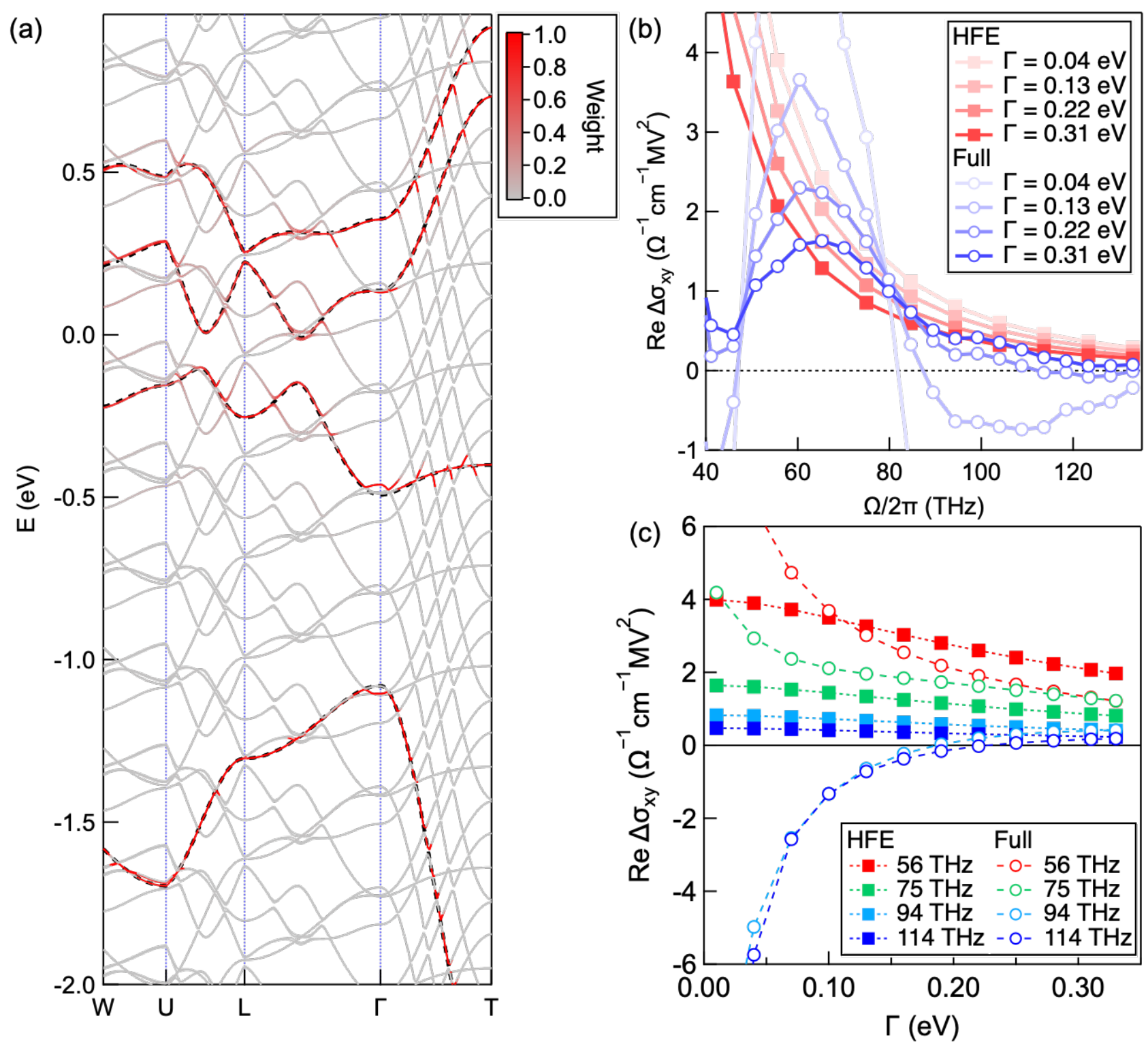

**Figure 16** (a) Floquet quasi-energy band structure of $Co_3Sn_2S_2$ tight-binding model calculated with including up to $\pm 15$-photon sectors for driving frequency $\Omega = 75$ THz. The electric field strength of the driving CPL is 1.0 MV/cm. Dashed lines indicate the non-driven band structure. (b) Driving frequency dependence of the AHC induced by the CPL with $E = 1.0\ \mathrm{MV/cm}$ with various energy broadening, $\Gamma$. Plotted in red (blue) are those calculated by high-frequency expansion (full Floquet) method. (c) The CPL-induced AHC as a function of $\Gamma$, for the calculations by the high-frequency expansion and the full Floquet methods.

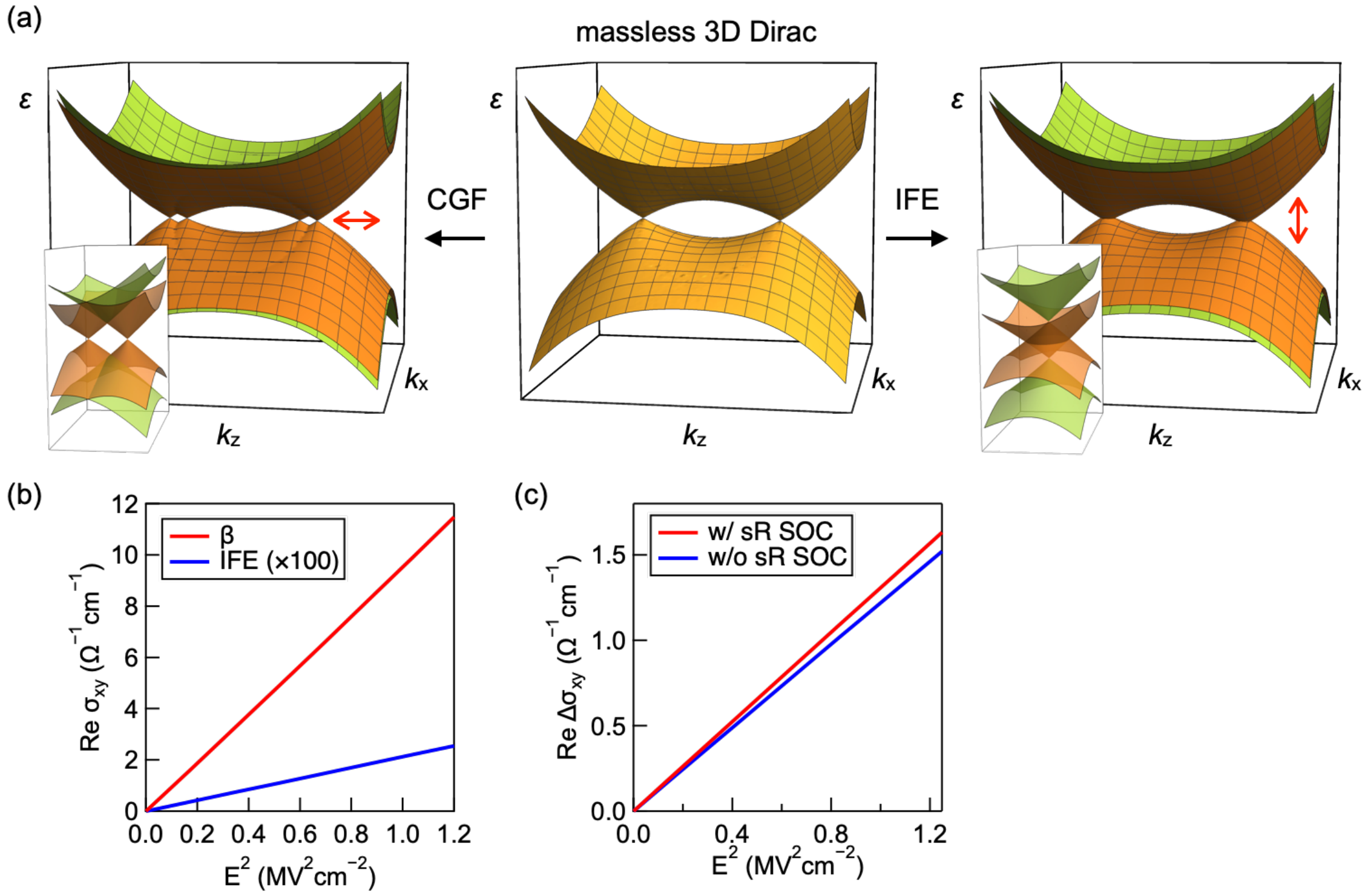

**Figure 17** (a) The band structures showing how the CGF and IFE modulate the massless 3D Dirac state. The insets are enlarged view around the initial DP. (b) AHC in the effective continuum model as a function of $E^2$ induced by the IFE (blue line) and the band splitting by $\beta$ (red line). (c) CPL-induced AHC calculated based on the effective tight-binding model of $Co_3Sn_2S_2$ as a function of $E^2$, with (red line) and without the staggered Rashba SOC (red line), respectively.